\documentclass[a4paper,12pt]{article}
\usepackage{graphicx,amssymb,amstext,amsmath,txfonts}
\usepackage{xcolor, yfonts, tcolorbox}
\usepackage{floatflt}
\usepackage{comment}
\usepackage{footnote}
\usepackage{float}
\usepackage{esint}
\usepackage{subcaption}
\usepackage{mathtools}
\usepackage{mathabx}
\usepackage{commath}
\usepackage{array}
\usepackage{boldline}
\usepackage{multirow}
\usepackage{braket}
\usepackage{comment}
\usepackage{arydshln}
\usepackage{bigdelim}
\usepackage{hyperref}
\usepackage{simpler-wick}
\usepackage[utf8]{inputenc}
\usepackage[style=phys,hyperref=true,articletitle=true,biblabel=brackets,chaptertitle=false]{biblatex}
\definecolor{cbl}{rgb}{0,0,1}                

\newcommand{\bc}{\begin{center}}
\newcommand{\ec}{\end{center}}
\def\ba#1{\begin{array}{#1}\displaystyle}
\newcommand{\ea}{\end{array}}

\newcommand{\beq}{\begin{equation}}
\newcommand{\eeq}{\end{equation}}
\newcommand{\beqa}{\begin{eqnarray}}
\newcommand{\eeqa}{\end{eqnarray}}

\newcommand{\bi}{\begin{itemize}}
\newcommand{\ei}{\end{itemize}}
\newcommand{\TTb}{\mathrm{T\bar{T}}}

\newcommand{\Tr}{{\rm Tr}}

\def \be {\begin{equation}} 
\def \ee {\end{equation}}

\def \la {\langle} 
\def \ra {\rangle}  

\begin{document}
\begin{titlepage}
\vspace{0.2cm}
\begin{center}

{\large{\bf{Generalised Hydrodynamics of $\TTb$-Deformed\\  Integrable Quantum Field Theories}}}

\vspace{0.8cm} 
{\large Riccardo Travaglino{\Large $^{{\color{teal}\star}}$}, Michele Mazzoni{\Large $^{{\color{blue}\spadesuit}}$} and Olalla A. Castro-Alvaredo{\Large $^{\LARGE\color{purple}\varheartsuit}$}}

\vspace{0.8cm}
{
{\Large $^{\color{teal} \star}$}  SISSA, via Bonomea 265, 34136 Trieste, Italy\\
\medskip

{\Large $^{{{\color{blue}\spadesuit}}{{\color{purple}\varheartsuit}}}$} Department of Mathematics, City, University of London, \\ 10 Northampton Square EC1V 0HB, UK\\
\medskip

}
\end{center}

\medskip
\medskip
\medskip
\medskip
\noindent
In this paper we evaluate the averages of conserved densities and currents associated to charges of generic spin in (1+1)-dimensional massive integrable Quantum Field Theories perturbed by the irrelevant $\TTb$ operator. By making use of the Thermodynamic Bethe Ansatz approach and of the theory of Generalised Hydrodynamics, we study the non-equilibrium steady state averages of conserved densities and  currents in a partitioning protocol. We show that in particular limits, averages can be evaluated exactly in terms of quantities known from the unperturbed theory. In the massless limit we recover known results for the energy and momentum currents and generalise those to any higher spin conserved quantities. We extend some of our results to perturbations of the generalised $\TTb$ type. For the massive free fermion theory, we find an analytic expression for the effective inverse temperature after at $\TTb$ perturbation in terms of the bare inverse temperature by making use of Lambert's $W$ function. 

\noindent 
\medskip
\medskip
\medskip
\medskip

\noindent {\bfseries Keywords:} $\TTb$ deformations, Generalized Hydrodynamics, Integrable Quantum Field Theory, Out-of-Equilibrium Systems, Generalized Gibbs Ensemble.

\vfill
\noindent 
{\Large $^{\color{teal} \star}$} rtravagl@sissa.it\\
{\Large $^{{{\color{blue}\spadesuit}}}$} michele.mazzoni.2@city.ac.uk\\
{\Large $^{{{\color{purple}\varheartsuit}}}$} o.castro-alvaredo@city.ac.uk

\hfill \today

\end{titlepage}
\tableofcontents
\section{Introduction}

In recent years there has been a lot of interest in the study of integrable quantum field theories (IQFTs) perturbed by a very particular irrelevant operator, namely  $\TTb$, which in $1+1$ dimensions is constructed starting from the holomorphic and anti-holomorphic components of the stress-energy tensor. This type of perturbation gives rise to theories that have many interesting properties. They are fundamentally non-local in nature, a feature that is often referred to as ``lack of UV completion" or ``UV fragility" \cite{Dubovsky:2012wk, Dubovsky:2013ira, Dubovsky:2017cnj}, in other words, the UV limit is not a conformal field theory (CFT). The introduction of the operator $\TTb$ and the study of some of its fundamental properties, notably the formula for its vacuum expectation value $\la {\TTb} \ra =- \la \Theta\ra^2$, where $\Theta$ is the trace of the stress-energy tensor, go back to the work \cite{Zamolodchikov:2004ce}. The works \cite{Smirnov:2016lqw,Cavaglia:2016oda} had the greatest influence in highlighting the special role of generalised $\TTb$-perturbed theories in the context of integrability. It was shown in \cite{Smirnov:2016lqw} that, under a generalised $\TTb$ perturbation, an IQFT remains integrable, in the sense that any scattering event factorises into two-body scattering processes and there is no particle production. {\iffalse\color{red}\fi This also means that the conserved charges of the original IQFT are deformed in such a way that they remain conserved in the perturbed model }.  The most recent understanding of this statement is that in fact,  generalised $\TTb$ perturbations are even more general than suggested in \cite{Smirnov:2016lqw}. In particular, they can include all local and quasi-local charges of the original unperturbed theory \cite{Doyon:2021tzy}. This makes it possible for instance to generate IQFTs that have a restricted set of local charges (ie. charges of just some odd spins) by a generalised $\TTb$-perturbation of a theory that has a different set of local charges (see ie. the recent work \cite{hotrep} where it is argued that the Lee-Yang model may be seen as a generalised $\TTb$-perturbation of the Ising field theory).

Considering a massive theory, at the $S$-matrix level the action of this deformation is simply given by a multiplicative CDD factor \cite{Castillejo:1955ed}. That is, if $S_{ab}^{cd}(\vartheta)$ is the two-body scattering amplitude of the process $a+b\mapsto c+d$ where the indices are particle quantum numbers, then the new scattering matrix is
\begin{equation}
    S_{ab}^{cd}(\vartheta) \longrightarrow e^{-i \delta_{ab}(\vartheta)} S_{ab}^{cd}(\vartheta)\qquad \mathrm{with} \qquad \delta_{ij}(\vartheta) = \sum_{s\in \mathcal{S}} \alpha_s m^s_a m^s_b \sinh(s\vartheta),
    \label{eq:ttbarSmatrix}
\end{equation}
 where $\alpha_s$ is a parameter of dimension $[M]^{-2s}$ which characterises the strength of the coupling and $m_a, m_b$ are the particle masses. The set $\mathcal{S}$ is typically that of the spins of local conserved charges. The term with $s=1$ corresponds to the standard $\TTb$ deformation, while higher spin terms correspond to generalised $\TTb$ deformations, or $\TTb_s$ deformations, which were also shown to exhibit an analogous form of solvability in \cite{Smirnov:2016lqw, Conti:2019dxg}. If the underlying theory is conformal, the factor  $\delta_{ab}(\vartheta)$ is instead introduced as an interaction term between the massless right and left movers of the theory, $\delta_{ab}(\vartheta)= \sum_s \alpha_s M^s_a M^s_b e^{s\vartheta}$, where the factors $M_a^s, M_b^s$ set the energy scale of the massless TBA equations \cite{ZAMOLODCHIKOV1991524,ZAMOLODCHIKOVcoset, bazhanov1996integrable}. See Section \ref{sec:tbadressing} for further discussion.
 
Generalised $\TTb$-perturbed theories are not only solvable, but many physical quantities can be explicitly related to their counterparts in the unperturbed theory. As we have seen, this is the case for the $S$-matrix, but also for thermodynamic quantities such as the free energy and the ground state energy \cite{Cavaglia:2016oda}, and, as shown very recently, for the matrix elements of local and twist fields (form factors) \cite{longpaper,PRL,ourentropy,theirentropy}. The correlation functions of primary operators in $\TTb$-deformed (1+1)-dimensional CFTs were also obtained in \cite{he2020correlation} as first order perturbations of the corresponding undeformed quantities. However, until recently it was not known how the solvability of these models reflects on their out-of-equilibrium dynamics. Building on the pioneering work \cite{Cavaglia:2016oda}, and subsequent results where this analysis has been extended and refined \cite{Conti:2019dxg,Hernandez-Chifflet:2019sua,Camilo:2021gro,Cordova:2021fnr,LeClair:2021opx,LeClair:2021wfd,Ahn:2022pia}, the works \cite{DavidTTb, MedenjakLong,MedenjakShort} studied the energy and momentum currents in $\TTb$-perturbed CFTs in a typical out-of-equilibrium protocol. This {\it partitioning protocol} is characterised by a scale-invariant initial state where two halves of a quantum system described by CFT are thermalised at different temperatures $T_R, T_L$ and then let to evolve for a long time until the system reaches a non-equilibrium steady state (NESS).  The main finding of \cite{MedenjakLong,MedenjakShort} consisted of showing that the known results for the currents in unperturbed CFT \cite{bernard2012energy,Bernard:2013aru,bernard2016conformal} are modified in the presence of a $\TTb$ perturbation in such a way that the currents are no longer of the form $f(T_L)-f(T_R)$ for some known function $f(x)$. This was also observed in \cite{DavidTTb} albeit only at first order in perturbation theory in the $\TTb$ coupling. The upshot is that, for $\TTb$-perturbed CFTs, there is instead a ``coupling" between the right and left temperatures, which reflects the new non-trivial interaction between the right- and left-moving massless excitations of the theory. The separation into functions of $T_R$ and $T_L$ does not generally extend to gapped theories (even in the absence of the $\TTb$ perturbation), however it is recovered in their massless limit, a property that extends to higher spin currents too, as recently shown \cite{MazzoniUnpublished,Mazzoni}. {\iffalse\color{red}\fi Although the focus of this work is on partitioning protocols in which the two sides are described by the same theory, the case of two halves described by different systems, such as CFTs with different central charges or spin chains with different values of a coupling constant, was studied in a series of works \cite{PhysRevB.98.075421,PhysRevB.93.205121,Bernard2014NonequilibriumCF}.}

In this paper, we approach the problem of out-of-equilibrium deformed theories from the viewpoint of Generalised Hydrodynamics (GHD) \cite{original_ghd,bertini_ghd}, a leading approach to computing large-scale dynamics of integrable models (see \cite{doyon_GHD_review,ESSLER_ghd_review} for reviews). As the terminology indicates, it is an approach based on hydrodynamic principles, therefore describes the physics of emergent behaviours in many-body quantum systems. As in classical hydrodynamics, GHD emerges naturally from local entropy maximisation over mesoscopic scale fluid cells containing a large number of quasiparticles. For IQFTs, it is well known that the TBA \cite{Zamolodchikov:1989cf} is the optimal framework which allows to obtain the thermodynamics, i.e. the maximal entropy states, of the (euclidean) field theory defined on a torus and treated in the $S$-matrix formulation. 
This is however not the full story, since it is well known that integrable systems do not thermalise in a standard sense: that is, the long time dynamics of some subsystem of length $L$ does not relax to a Gibbs state \cite{Kinoshita2006AQN}, in the sense of
\begin{equation}
    \lim_{t\to \infty} \lim_{L\to \infty}\braket{\hat{O}} = \Tr{\left[\rho_{GE}\hat{O}\right]}\,.
\end{equation}
This is due to the presence of an infinite tower of conserved charges, which have to be considered in the assumption 
that the time evolution will lead to a state which retains the minimal amount of information on the initial state \cite{ESSLER_ghd_review}. The system then equilibrates  to a Generalised Gibbs Ensemble (GGE) \cite{GGE_definition}:
\begin{equation}
    \rho_{GGE} \propto e^{-\sum_s \beta_s Q_s}\,,
\end{equation}
where the operators $Q_s$ form the full set of conserved changes (as mentioned earlier, this can include non-local charges \cite{enej}). This change of ensemble leads to a natural modification of the standard (thermal) TBA equations \cite{Fioretto:2009yq,Mossel:2012vp}, as we shall see below. We can therefore say that GHD is a ``local version", in the (hydrodynamic) sense of fluid cells, of the TBA approach, in which the TBA equations are suitably modified to take into account the GGE.
GHD can therefore be used to study inhomogeneous and non-equilibrium phenomena involving IQFTs. In particular, in its simplest form, it can be used to evaluate (usually numerically) the averages of conserved currents and densities of any spin. {\iffalse\color{red}\fi Given a local charge $Q_s$ we can express it in terms of a local density $q_s$}, $Q_s=\int dx q_s(x,t)$, from which we can find the associated current by a continuity equation $\partial_t q_s(x,t) = -\partial_x j_s(x,t)$. The works \cite{original_ghd,bertini_ghd} provided a prescription for computing averages of these local densities for quantum integrable models. This is the prescription that we use in this work. 

Instead of considering a CFT as it was done in \cite{MedenjakLong,MedenjakShort}, we  start from the $T\bar{T}$ deformation of a massive IQFT.  We recover CFT results in the massless limit, including those of \cite{MedenjakLong,MedenjakShort}. {\iffalse\color{red}\fi In the massless limit}, we then go on to generalise these results to higher spin currents and densities, obtaining the new formulae
\beq
\begin{split}
       \qquad {\texttt q}_{\pm s}^\alpha &= G(s)c_{LR}\left(\left(\hat{T}_L^{s+1} \pm\hat{T}_R^{s+1}\right) - \frac{\alpha \pi c}{6}(\hat{T}_L^{s+1} \hat{T}_R^2 \pm\hat{T}_R^{s+1}\hat{T}_L^2) \right), \qquad \\
       \qquad  {\texttt j}_{\pm s}^\alpha &= G(s)c_{LR}\left(\left(\hat{T}_L^{s+1} \mp\hat{T}_R^{s+1}\right) + \frac{\alpha \pi c}{6} (\hat{T}_L^{s+1} \hat{T}_R^2 \mp\hat{T}_R^{s+1}\hat{T}_L^2) \right).\qquad 
         \end{split}
         \label{critical}
\eeq
Here, ${\texttt q}_{\pm s}^\alpha$ represent spin $s$ density averages of even/odd charges, that is charges whose one-particle eigenvalues are even/odd functions of the rapidity, in the NESS reached after a partitioning protocol, and similarly for the currents ${\texttt j}_{\pm s}^\alpha$. {\iffalse\color{red}\fi Even charges and currents will be labeled by positive spin index +s, while odd charges and currents by negative spin -s. We stress that this terminology does not refer to the spin itself but to the parity of the charge eigenvalues as functions of $\vartheta$}.
The {\iffalse\color{red}\fi parameter $\alpha$ appearing in \eqref{critical}} refers to the coupling $\alpha:=\alpha_1$ in (\ref{eq:ttbarSmatrix}), that is, these are the averages corresponding to a generic CFT perturbed by $\TTb$ only. The generalised temperatures $\hat{T}_R, \hat{T}_L$ are related to the right/left temperatures in the original baths. They are in fact effective versions of those, as they reduce to the ``bare" temperatures $T_{R,L}$ when $\alpha=0$. For generic $\alpha$ we have the non-trivial identity $\hat{T}_{R,L}=T_{R,L}(1-\alpha {T}_{R,L} E_0^\alpha)^{-1}$, where $E_0^\alpha$ is the ground state energy of the deformed theory. {The quantity \iffalse\color{red}\fi $G(s)$ in \eqref{critical} is a theory-dependent function of the spin which was introduced and computed exactly for free theories in \cite{MazzoniUnpublished}. For interacting theories it is only known for $s=1$ ($G(1)=\frac{\pi c}{12}$) \cite{Zamolodchikov:1989cf}}. Finally, $c_{LR}$ is a function of the effective temperatures such that $c_{LR} = 1$ for $\alpha=0$. 
 Besides the formulae (\ref{critical}), which are valid at critical points, we have found more general relations between the perturbed and unperturbed currents and densities which are also valid away from criticality, both for partitioning protocol and at equilibrium. 
 
\medskip

The paper is structured as follows: in Section \ref{sec:tbadressing} we review the TBA equations for $\TTb$-deformed IQFTs and show how the $\TTb$ deformation of the $S$-matrix affects the equations that describe the dressing of the single particle eigenvalues of conserved quantities. In Section \ref{sec:freefermion} we study the free fermion theory, introducing techniques which will be fully developed in the next section. In Section \ref{sec:mainresults} we present our main results, including universal formulae for the average densities and currents of higher spin quantitities in generic CFT. 
 In Section \ref{scaling} we discuss some general properties of the TBA scaling function, such as its monotonicity as a function of $m, \alpha$ and $\beta$.  In Section \ref{numerics} we present numerical results and discuss their physical implications. We conclude in Section \ref{conclusion}. Various extensions of the work are presented in the Appendices. In Appendix \ref{appA} we consider theories with many particles. In Appendix \ref{appB} we analyse the case of more general $\TTb$ perturbations, namely those associated with a spin $s$ conserved change. In Appendix \ref{appC} we present a CFT derivation of some of our results. In Appendix \ref{appD} we
derive an equilibrium small mass expansion of the effective inverse temperature for the $\TTb$-perturbed massive free fermion.

\section{$\TTb$-Deformed TBA Equations and Dressing}
\label{sec:tbadressing}

It is well known that the Thermodynamic Bethe Ansatz \cite{Zamolodchikov:1989cf} provides the theoretical framework to study the thermodynamics of an IQFT. It essentially reduces the problem of evaluating the partition function of the system to the problem of solving a set of coupled integral equations for the energy of the  elementary excitations of the system, the TBA equations.
For systems with a single massive particle of mass $m$ there is only one equation, which takes the form:
\begin{equation}
    \varepsilon(\vartheta)= \nu(\vartheta) - (\varphi* L)(\vartheta),\qquad \mathrm{with} \qquad \hspace{0.3cm} L(\vartheta) := \ln(1+e^{-\varepsilon(\vartheta)}),
\end{equation}
where $*$ represents the convolution, 
\beq 
(a* b)(\vartheta):=\frac{1}{2\pi}\int_{-\infty}^\infty d\beta \, a(\vartheta-\beta)b(\beta)\,.
\label{convo}
\eeq 
The scattering kernel $\varphi(\vartheta)$ is {\iffalse\color{red}\fi proportional to} the logarithmic derivative of the $S$-matrix $\varphi(\vartheta):=-i \frac{S'(\vartheta)}{S(\vartheta)}$, $\varepsilon(\vartheta)$ is the pseudoenergy and $\nu(\vartheta)$ is the driving term. 
In a GGE the driving term can have a very general form, resulting from the inclusion of the one-particle eigenvalues of local and quasi-local charges of any conserved spin: 
\begin{equation}
    \nu(\vartheta)=\sum_{s} \beta_s m^s \cosh(s\vartheta) + \sum_{s'} \gamma_{s'} m^{s'} \sinh(s'\vartheta).
\end{equation}
 The one-particle eigenvalues $h_s(\vartheta)$, with $s=\{\pm 1, \pm2,...\}$, can be obtained by  differentiating the driving term with respect to the generalised thermodynamic potentials, 
\beq{
h_s(\vartheta) = \frac{\partial \nu(\vartheta)}{\partial \beta_s}=m^s\cosh(s\vartheta), \qquad h_{-s}(\vartheta) = \frac{\partial \nu(\vartheta)}{\partial \gamma_s}=m^s\sinh(s\vartheta). }
\eeq
In particular, the energy and momentum eigenvalues enter many important formulae, and for those it is common to use the notations $h_1(\vartheta)=E(\vartheta)=m\cosh\vartheta$ and $h_{-1}(\vartheta)=P(\vartheta)=m\sinh\vartheta$.

We can now introduce the averages of currents and densities as
\begin{equation}
       {\texttt q}_s:= \int \frac{d\vartheta}{2\pi} E(\vartheta) n(\vartheta) h_s^{dr}(\vartheta), \qquad \mathrm{and}\qquad
        {\texttt j}_s:= \int \frac{d\vartheta}{2\pi} P(\vartheta) n(\vartheta) h_s^{dr}(\vartheta)\,, \label{qjdef}
\end{equation}
where $n(\vartheta)=(1+e^{\varepsilon(\vartheta)})^{-1}$ is the occupation function. The dressed eigenvalue $h_s^{\rm dr}(\vartheta)$ is 
\begin{equation}
   h_s^{\rm dr}(\vartheta) = \frac{\partial \varepsilon(\vartheta)}{\partial \beta_s} = h_s(\vartheta) + (\varphi * g_s)(\vartheta),\qquad \mathrm{with}\qquad g_s(\vartheta)=n(\vartheta)h_s^{\rm dr}(\vartheta)\,.
   \label{dressing1}
\end{equation}
The dressing equation describes how the eigenvalue of a given charge and quasiparticle is modified by interaction with other quasiparticles, the interaction being encoded in the scattering kernel\footnote{The fact that the dressing operation is defined by differentiating the TBA equations w.r.t. the Lagrange multipliers $\beta_i$ is a formal construction, meaning that the dressing equations are meaningful even in cases when those parameters are zero. For instance, in this paper we will mostly consider the TBA with a thermal driving term. Nonetheless, one can dress higher charges and compute their averages also in this case.}. Clearly, for free theories where $\varphi(\vartheta)=0$, the dressing operation is trivial and $\varepsilon(\vartheta)=\nu(\vartheta)$.

The particularly simple way in which the $S$-matrix is modified by the addition of the $\TTb$ deformation leads to a simple modification of the TBA equations through the introduction of an additional term in the scattering kernel, that is, from (\ref{eq:ttbarSmatrix}) we have
\begin{equation}
    \varphi_{ab}(\vartheta) \mapsto  \varphi_{ab}(\vartheta) - \alpha m_a m_b \cosh\vartheta\,.
\end{equation}
For a theory with a single particle, we will call $\varphi^\alpha(\vartheta)$ the deformed kernel and $\varphi^0(\vartheta)$ the original kernel. Since all TBA quantities will now depend on $\alpha$ we will also adopt notations $\varepsilon^\alpha(\vartheta)$, $n^\alpha(\vartheta)$ and $L^\alpha(\vartheta)$ for the standard TBA functions. Then the convolution acts as
\begin{equation}
    (\varphi^\alpha * L^\alpha)(\vartheta)=(\varphi^0 * L^\alpha)(\vartheta) -\alpha m^2 (\cosh * L^\alpha)(\vartheta),
\end{equation}
with
\begin{eqnarray}
     m (\cosh * L^\alpha)(\vartheta)=\frac{m}{2\pi}\int_{-\infty}^\infty \cosh(\vartheta-\vartheta') L^\alpha(\vartheta') d\vartheta'=-E_0^\alpha \cosh\vartheta+ P_0^\alpha \sinh\vartheta \,,
\end{eqnarray}
where
\begin{equation}
    E_0^\alpha=-\frac{m}{2\pi} \int_{-\infty}^\infty \cosh \vartheta L^\alpha(\vartheta) d\vartheta\,,\hspace{1cm} P_0^\alpha=-\frac{m}{2\pi} \int_{-\infty}^\infty \sinh \vartheta L^\alpha(\vartheta) d\vartheta\,,
    \label{EP}
\end{equation}
are the ground state energy and total momentum. We can then write the equilibrium TBA equation at inverse temperature $\beta$ as
\begin{equation}
\label{eq:generalformofperturbedtba}
    \varepsilon^\alpha(\vartheta) =(\beta-\alpha E_0^\alpha)m \cosh\vartheta +   \alpha m P_0^\alpha \sinh\vartheta - (\varphi^0 * L^\alpha)(\vartheta)\,.
\end{equation} 
Here we have considered a thermal driving term $\nu(\vartheta)=m\beta \cosh\vartheta$.
At equilibrium $P_0^\alpha=0$, since $L^\alpha(\vartheta)$ is an even function, therefore in this situation the effect of the perturbation is akin to a redefinition of the inverse temperature, $\beta \to \beta-\alpha E_0^\alpha $\footnote{If both the driving term and the perturbation are chosen more generally, the overall effect is that of replacing the TBA equation for a particular GGE with the TBA equation for a different GGE, in the sense that some of the generalised inverse temperatures are modified \cite{Hernandez-Chifflet:2019sua}. We will consider some of these situations in Appendix \ref{appB}.}. {This fact is physically very interesting and suggests that the effect of a $\TTb$ deformation is that of either reducing or increasing the temperature ``felt" by the system. There are different viewpoints on why this might be the case, some based on integrability \cite{Cardy:2020olv}
some on the connection between $\TTb$ perturbations and coupling to $JT$ gravity \cite{Dubovsky:2018bmo,Conti:2018jho, Conti:2018tca, Dubovsky:2023lza}  but they all boil down to relating finite temperature to finite volume in the mirror TBA picture, and then identifying $m\alpha$ as a new length scale. Following \cite{Cardy:2020olv} a $\TTb$ perturbation transforms local degrees of freedom into extended degrees of freedom. Effectively, for $\alpha < 0$ particles acquire a finite width and so the overall available volume is decreased. The reverse occurs for $\alpha>0$, in which case particles acquire an effective negative width and the overall available volume is increased. In the gravity picture, the effect of the gravity field is a modification of the temperature/system size as discussed already in very early works such as \cite{Lut}.}

We keep the dependence on $P_0^\alpha$ below, since this will be useful when we consider out-of-equilibrium situations later on. For $P_0^\alpha \neq 0$ the $\TTb$ perturbation introduces a state-dependent redefinition of the temperature and a Lorenz boost. 
It is interesting to note that the expression above can be rewritten as:
\begin{equation}
\label{generaltba}
     \varepsilon^\alpha(\vartheta) =\left(\beta-\alpha (E_0^\alpha -  P_0^\alpha)\right)\frac{m}{2} e^\vartheta +   \left(\beta-\alpha (E_0^\alpha + P_0^\alpha)\right)\frac{m}{2} e^{-\vartheta} - \varphi^0 * L^\alpha(\vartheta),
\end{equation}
from which it is immediate to take the massless limit \cite{Zamolodchikov:1992zr}: letting $m\to 0$ and $\vartheta \mapsto \vartheta_0 + {\vartheta}$, with $\vartheta_0 \rightarrow \infty$, we can define a new finite non-zero energy scale $M:= m e^{\vartheta_0}$ and obtain the TBA equations for the CFT right/left ($\pm$) movers:
\begin{equation}
\label{eq:perturbedcft}
    \varepsilon^{\alpha}_\pm(\vartheta) =\frac{M}{2}\left(\beta-\alpha (E_0^\alpha \mp P_0^\alpha)\right) e^{\pm \vartheta}   - (\varphi^0 * L_{\pm}^\alpha)(\vartheta)\,,
\end{equation}
which is related to the massless Bazhanov-Lukyanov-Zamolodchikov TBA equation \cite{bazhanov1996integrable}. This derivation is slightly different from \cite{MedenjakShort,Cavaglia:2016oda}, where an explicit interaction between right and left movers is introduced, {\iffalse\color{red}\fi but the two TBA equations obtained are equivalent}. This formulation highlights the fact that the two TBA equations remain separated, and the interaction is purely given by a ``mean field" effect through a the total energy and momentum.

From the definitions above it is easy to show that (formally) $\frac{\partial E_0^\alpha}{\partial \beta_s} = {\texttt q}_s^\alpha$ and $\frac{\partial P_0^\alpha}{\partial \beta_s} = {\texttt j}_s^\alpha$ as defined by (\ref{qjdef}). It is then immediate to obtain the dressing equation for a $\TTb$ perturbed theory,
\beq 
\label{eq:dressing}
h_s^{\rm{dr},\alpha}(\vartheta)=h_s(\vartheta)-\alpha E(\vartheta) {\texttt q}^\alpha_s + \alpha P(\vartheta) {\texttt j}^\alpha_s +(\varphi_0 * g_s^\alpha)(\vartheta)\,,
\eeq 
to be compared with (\ref{dressing1}).
The fact that the dressing equation is modified by terms which are proportional to the average densities and currents, that is, the same quantities we want to compute, is crucial in order to find formulae for ${\texttt q}_s^\alpha$ and ${\texttt j}_s^\alpha$ in terms of their underformed counterparts ${\texttt q}_s^0$ and ${\texttt j}_s^0$. {\iffalse\color{red}\fi Understanding the relation between perturbed and unperturbed quantities will be the object of much of this paper}. Note that, just as for the total momentum above, which is zero at equilibrium, also the (even) currents vanish at equilibrium (that is, ${\texttt j}_s^\alpha=0$). Nonetheless we write these explicitly here, since it will make it easier to generalise the equations to out-of-equilibrium/GGE situations. It is easy to extend this analysis for theories with a multi-particle spectrum, see Appendix \ref{appA}. 

Finally, a few words on the partitioning protocol, the only truly out-of-equilibrium protocol we will consider in this paper. This protocol was also the focus of \cite{Bernard:2013aru,bernard2016conformal,original_ghd,bertini_ghd}. As outlined in the introduction, we consider two subsystems thermalised at temperatures $T_L$ and $T_R$. Taking this as our initial condition and joining the two subsystems at $x=t=0$, the large-time evolution leads to a NESS developing around $x=0$,  with non-trivial currents present in the system. 
The initial state can be represented in terms of the occupation function as
\begin{equation}
    n^\alpha(\vartheta,x,0) = \begin{cases}
        n^\alpha_L(\vartheta) = n^\alpha(\vartheta)\Bigr|_{\{\beta_L^s\}} \text{ for } x<0 \\
         n^\alpha_R(\vartheta) = n^\alpha(\vartheta)\Bigr|_{\{\beta_R^s\}} \text{ for } x>0
    \end{cases}\,.
\end{equation}
The general solution for the current at the contact point between the two halves was found in \cite{original_ghd,bertini_ghd}:
\begin{equation}
\label{n_partitioning}
    n^\alpha(\vartheta) = n^\alpha_L(\vartheta) \Theta(\vartheta-\vartheta^\alpha_\star) + n^\alpha_R(\vartheta) \Theta(\vartheta^\alpha_\star-\vartheta)\,,
\end{equation}
where the value $\vartheta_\star^\alpha$ is the solution\footnote{This is unique if the effective velocity is a monotonic function of the rapidity, which is the case in all the theories we will consider in this work. {\iffalse\color{red}\fi The monotonicity property is generally lost only in rather exotic models, such as Zamolodchikov's staircase model \cite{zamolodchikov2006resonance}, for which a more sophisticated treatment is required \cite{Mazzoni}.}} to $v^{\rm eff,\alpha}(\vartheta^\alpha_\star)=0$, and the effective velocity is defined as 
\beq 
v^{\rm eff,\alpha}(\vartheta)=\frac{P^{\rm dr,\alpha}(\vartheta)}{E^{\rm dr, \alpha}(\vartheta)}\,.
\label{veff1}
\eeq 
Therefore we see that the solution is determined entirely by the right/left equilibrium solutions and by the value of $\vartheta^\alpha_\star$. 
\section{The Free Fermion}
\label{sec:freefermion}
Free theories provide an ideal example where the formulae presented above can be analysed in more detail and it is possible to obtain exact analytic solutions. In this case $\varphi^0(\vartheta)=0$ and the TBA equations can be solved exactly, even for the $\TTb$-perturbed theory. At equilibrium, we have 
\beq 
\label{eq:tbawithenergy}
\varepsilon^\alpha(\vartheta)=(\beta -\alpha E_0^\alpha) m\cosh\vartheta\,,
\eeq 
with 
\beq 
\label{eq:groundstateenergy}
E_0^\alpha=-\frac{m}{2\pi} \int_{-\infty}^\infty \cosh\vartheta \log(1+e^{-(\beta  -\alpha E_0^\alpha )m\cosh\vartheta}) d\vartheta\,.
\eeq 
As discussed in \cite{Smirnov:2016lqw,Cavaglia:2016oda}, the ground state energy of the $\TTb$-perturbed theory admits an expression which depends non-linearly on the undeformed ground state energy. This relation is encoded in the fact that the deformed ground state energy satisfies the inviscid Burgers' equation. For the free fermion, equation (\ref{eq:groundstateenergy})  can be solved exactly by expanding the logarithm and then using Bessel functions, generalising the free fermion treatment presented in \cite{KLASSEN1991635}. We observe that for $\beta-\alpha E_0^\alpha > 0 $ we can expand the logarithm, and introduce the modified Bessel function of the second kind:
\beq
K_a(z) = \int_0^\infty e^{-z\cosh t}\cosh{(a t)}dt,
\eeq
so as to obtain:
\beqa 
E_0^\alpha&=&\frac{m}{2\pi} \sum_{n=1}^\infty \frac{(-1)^n}{n}\int_{-\infty}^\infty \cosh\vartheta e^{-n(\beta  -\alpha E_0^\alpha )m\cosh\vartheta} d\vartheta\nonumber\\
&=&\frac{m}{\pi} \sum_{n=1}^\infty \frac{(-1)^n}{n} K_1(n (\beta  -\alpha E_0^\alpha )m)\approx \frac{m}{\pi} \sum_{n=1}^\infty \frac{(-1)^n}{n^2 (\beta  -\alpha E_0^\alpha )m} \nonumber\\
&=& \frac{1}{\pi (\beta  -\alpha E_0^\alpha )}\sum_{n=1}^\infty \frac{(-1)^n}{n^2}=- \frac{\pi}{12(\beta  -\alpha E_0^\alpha )}\nonumber\\
&=&- \frac{\pi c}{6 (\beta  -\alpha E_0^\alpha )}\qquad \mathrm{for} \qquad m\ll 1 \label{ole}\,,
\eeqa 
where we have used the expansion of the Bessel function for small argument, $K_1(z) \sim \frac{1}{z}$, and introduced the central charge of the free fermion $c=1/2$ so as to recover an expression which in fact holds for generic CFT. We observe that for $\alpha=0$ we recover the known formula $E_0^0:=-\frac{\pi c}{6\beta}$. 
From \eqref{ole} we obtain a quadratic equation in $E_0^\alpha$
\begin{equation}
    \alpha (E_0^\alpha)^2 - \beta E_0^\alpha -\frac{\pi c}{6} = 0\,,
\end{equation}
which can be solved to:
\beq
E_0^\alpha=\frac{\beta}{2  \alpha }\left(1\pm \sqrt{1+ \frac{2\alpha \pi c}{3\beta^2}}\right)\,.
\label{E0}
\eeq 
Although we used the free fermion as our example, this expression is valid for any CFT, as shown for instance in \cite{Cavaglia:2016oda}.
We see that for $\alpha<0$ the energy can become complex and has a square root branch point. This is related to the famous Hagedorn transition \cite{Hagedorn:1965st}. In order to avoid this complication, we will limit ourselves to the $\alpha>0$ case. Moreover, of the two possible signs in (\ref{E0}) we will take only the negative sign, as it is the one for which the energy remains finite as $\beta \to \infty$, and for which the condition $\beta - \alpha E_0^\alpha > 0$ holds.
Introducing the scaling function $c^\alpha$ as $E^\alpha_0:=-\frac{\pi c^\alpha}{6\beta}$ we obtain precisely the same formula as in \cite{MedenjakShort,MedenjakLong}, (with the identification $\alpha=-\sigma/2$):
\beq
    c^\alpha = -\frac{3\beta^2}{\pi \alpha }\left(1 - \sqrt{1+ \frac{2\alpha \pi c}{3\beta^2}}\right)\,.
    \label{eq:c_function_medenjak}
\eeq
Interestingly, for $\TTb$-perturbed theories, the scaling function is $\beta$-dependent in the conformal limit. In fact, it depends on the only mass-independent dimensionless scale of the problem, namely the ratio $\beta^2/\alpha$. It is useful to introduce the effective inverse temperature 
\beq 
\label{eq:defBeta}
\hat{\beta}:=\beta-\alpha  E_0^\alpha\,,
\eeq 
which in CFT is given by:
\beq 
 \hat{\beta}= \frac{\beta}{2}\left(1+  \sqrt{1+ \frac{2\pi c\alpha}{3\beta^2}}\right)\,.
 \label{confdef}
\eeq 
Beyond the critical point, there is generally no analytic formula for the relationship between $\hat{\beta}$ and $\beta$. However, for the massive free fermion a formula can still be obtained (see Appendix~\ref{appD}).

Let us now consider the averages of currents and densities.  Using the dressing equation \eqref{eq:dressing} we can immediately compute the effective velocity of the theory 
\begin{equation}
\label{eq:effvelfreefermion}
    v^{\rm eff,\alpha}(\vartheta) = \frac{P^{\rm dr, \alpha}(\vartheta)}{E^{\rm dr,\alpha}(\vartheta)} = \frac{P(\vartheta) - \alpha {\texttt q}_{-1}^\alpha E(\vartheta)+\alpha {\texttt j}_{-1}^\alpha P(\vartheta)}{E(\vartheta) - \alpha {\texttt q}_1^\alpha E(\vartheta)+\alpha {\texttt j}_1^\alpha P(\vartheta)}\,.
\end{equation}
The crucial quantity in the partitioning protocol is the value $\vartheta^\alpha_\star$. This can be easily found from the equation above:
\begin{equation}
    v^{\rm eff,\alpha}(\vartheta^\alpha_\star) = 0 \Leftrightarrow \tanh(\vartheta^\alpha_\star) = \frac{\alpha {\texttt q}_{-1}^\alpha}{1+\alpha {\texttt j}_{-1}^\alpha}\,.
\end{equation}
Numerically, this can be used to find the currents and densities in a self-consistent fashion. However, if we are interested in an analytical result, this is only possible in either the  massless $m\to 0$ or the unperturbed $\alpha\rightarrow 0$  limits. In both cases \footnote{Regarding the $m\to 0$ limit, the statement is true only in a sense which will be clarified for the general case in section \ref{42}}
\beq
    \lim_{m \to 0} \vartheta^\alpha_\star = \lim_{\alpha \to 0} \vartheta^\alpha_\star= 0\,.
\eeq
Therefore, in the massless limit the value of the $\vartheta_\star^\alpha $ is precisely the same as for the free fermion without the perturbation. This makes the study of the partitioning protocol much easier. Indeed, consider the expressions for the currents and densities, which we expand by making use of \eqref{eq:dressing}:
\beqa
    {\texttt j}_s^\alpha = \int \frac{d\vartheta}{2\pi} P(\vartheta) n^\alpha(\vartheta) \left( h_s(\vartheta) - \alpha E(\vartheta) {\texttt q}_s^\alpha +\alpha P(\vartheta) {\texttt j}_s^\alpha  \right), \label{js_free}\\
    {\texttt q}_s^\alpha = \int \frac{d\vartheta}{2\pi} E(\vartheta) n^\alpha (\vartheta) \left( h_s(\vartheta) - \alpha  E(\vartheta) {\texttt q}_s^\alpha +\alpha P(\vartheta){\texttt j}_s^\alpha  \right).
    \label{qs_free}
\eeqa
It is possible to argue, more generally, (see Subsection~\ref{42}) that for high enough temperatures one can take $\vartheta_\star^\alpha \approx \vartheta_\star^0$. Under this approximation, the function $n^\alpha(\vartheta)$ is exactly the same as the one for the unperturbed theory up to the redefinition of temperature (\ref{eq:defBeta}):
\beq
n^\alpha(\vartheta)\bigr|_{(\beta_R,\, \beta_L)} =n^0(\vartheta)\bigr|_{(\hat\beta_R,\, \hat\beta_L)}.
\eeq
Therefore, the quantities \eqref{js_free}, \eqref{qs_free} can be expressed in terms of average densities and currents in the partitioning protocol of a free fermion at inverse temperatures $\hat{\beta}_R$ and $\hat{\beta}_L$. Let us denote those average densities and currents by $\hat{\texttt \j}_s^0, \hat{\texttt q}_s^0$, where the hat denotes dependence on the effective temperatures. We then obtain the equations
\begin{equation}
    {\texttt j}_s^\alpha = \hat{\texttt \j}_s^0 -\alpha {\texttt q}_s^\alpha \hat{\texttt q}_{-1}^0 +\alpha {\texttt j}_s^\alpha \hat{\texttt \j}_{-1}^0 \qquad \mathrm{and}\qquad 
                {\texttt q}_s^\alpha = \hat{\texttt q}_s^0 - \alpha {\texttt q}_s^\alpha \hat{\texttt q}_1^0 +\alpha {\texttt j}_s^\alpha \hat{\texttt \j}_1^0\,.
                \label{34}
\end{equation}
Assuming that the charges and currents of the unperturbed theory are known (they are known exactly for free theories, see \cite{Mazzoni,MazzoniUnpublished}) this is a system of equations for  ${\texttt j}_s^\alpha$ and ${\texttt q}_s^\alpha$ with solutions:
\beqa 
        {\texttt q}_s^\alpha &=& \frac{\hat{\texttt q}_s^0 +\alpha \hat{\texttt \j}_{1}^0 \hat{\texttt \j}_s^0 -\alpha \hat{\texttt \j}_{-1}^0 \hat{\texttt q}_s^0}
        {1 +\alpha (\hat{\texttt q}_1^0- \hat{\texttt \j}_{-1}^0) +\alpha^2( \hat{\texttt \j}_1^0\hat{\texttt q}_{-1}^0- \hat{\texttt \j}_{-1}^0 \hat{\texttt q}_1^0 ) }\,, \label{qs}\\  
        {\texttt j}_s^\alpha &=& \frac{\hat{\texttt \j}_s^0 +\alpha \hat{\texttt \j}_s^0 \hat{\texttt q}_1^0 -\alpha \hat{\texttt q}_{-1}^0 \hat{\texttt q}_s^0}{1 +\alpha (\hat{\texttt q}_1^0- \hat{\texttt \j}_{-1}^0) +\alpha^2( \hat{\texttt \j}_1^0\hat{\texttt q}_{-1}^0- \hat{\texttt \j}_{-1}^0 \hat{\texttt q}_1^0 ) } \label{js}\,.
\eeqa 
These formulae relate the average currents and densities in a $\TTb$-deformed fermion at given temperature(s) to those of an unperturbed free fermion at the effective temperature(s). The formulae are exact at equilibrium and hold also in the partitioning protocol with the aforementioned approximation $\vartheta_\star^\alpha \approx \vartheta_\star^0$. In particular, at equilibrium the formulae above can be further simplified since the currents associated to even charges vanish (whereas odd ones, like the momentum current, in general do not). Therefore, the equilibrium average densities simplify to:
\begin{equation}
    {\texttt q}_s^\alpha = \frac{\hat{\texttt q}_s^0 -\alpha \hat{\texttt \j}_{-1}^0 \hat{\texttt q}_s^0 }{1-\alpha \hat{\texttt \j}_{-1}^0 + \alpha \hat{\texttt q}_1^0 -\alpha^2 \hat{\texttt \j}_{-1}^0 \hat{\texttt q}_1^0} =  \frac{\hat{\texttt q}_s^0}{1 + \alpha \hat{\texttt q}_1^0}\,.
    \label{freef}
\end{equation}
Note however that even at equilibrium all solutions depend on $\hat{\beta}$, which is known analytically as a function of $\beta$ only at the critical point. 
Away from that, $\hat{\beta}$ has to be obtained through the solution of the inviscid Burgers' equation \cite{Smirnov:2016lqw,Cavaglia:2016oda}.

\section{Interacting Theories}
\label{sec:mainresults}
Consider now an interacting theory with a single particle spectrum. In this situation, the TBA equation is of the form \eqref{generaltba}. To be as general as possible, we will not specify for now whether we are in the equilibrium case or the partitioning protocol. We will start by revisiting the dressing operation. Following \cite{doyon_GHD_review}, we  can write the convolution term $(\varphi^0 * (n\, h^{\rm dr}))(\vartheta)$ by means of an integral operator $\mathbf{T}$ such that $(\varphi^0 * (n\, h^{\rm dr}))(\vartheta)=:({\bf T} n)h^{\rm dr}(\vartheta)$. Thus, in the undeformed theory equation (\ref{dressing1}) reads:
\beqa
    h_s^{\rm dr,0}(\vartheta) = (1 - \mathbf{T} n^0)^{-1}  h_s(\vartheta)\,.
    \label{eq:freedressing}
\eeqa
The above equation should be understood as a formal power series in $\mathbf{T}$:
\begin{equation}
    f(\mathbf{T}) = \sum_{n=0}^{\infty}\frac{f'(0)}{n!}\mathbf{T}^n\,,
\end{equation}
where the powers of the integral operator are interpreted as multiple convolutions.
Therefore we are identifying the dressing operation in the unperturbed theory with the action of the integral operator $(1 - \mathbf{T} n^0)^{-1}$ on the bare charge eigenvalues. The $\TTb$ deformation leads to the addition of two extra terms in the dressing equations, so that the same manipulation, applied to equation \eqref{eq:dressing}, yields:
\beq
    h^{\rm dr,\alpha}_s(\vartheta) = (1 - \mathbf{T} n^\alpha)^{-1} ( h_s(\vartheta) - \alpha {\texttt q}_s^\alpha E(\vartheta) + \alpha {\texttt j}_s^\alpha P(\vartheta))\,,
    \label{dressingT}
\eeq
which we rewrite conveniently as:
\beq
    h^{\rm dr, \alpha}_s(\vartheta) = \tilde{h}_s^\alpha(\vartheta) - \alpha {\texttt q}_s^\alpha \Tilde{E}^\alpha(\vartheta) + \alpha {\texttt j}_s^\alpha \Tilde{P}^\alpha(\vartheta)\,,
    \label{eq:TildeDressingEquation}
\eeq 
where we use linearity of ${\bf T}$ and defined 
tilded quantities as:
\begin{equation}
    \Tilde{A}^\alpha(\vartheta) := (1 - \mathbf{T} n^\alpha)^{-1} A(\vartheta)\,.
    \label{eq:tilde}
\end{equation}
We then obtain similar formulae as for the free case, now in terms of tilded quantities:
\begin{align}
    {\texttt q}_s^\alpha = \int \frac{d\vartheta}{2\pi} E(\vartheta) n^\alpha(\vartheta) \left( \Tilde{h}^\alpha_s(\vartheta) - \alpha \Tilde{E}^\alpha(\vartheta) {\texttt q}_s^\alpha +\alpha \Tilde{P}^\alpha(\vartheta) {\texttt j}_s^\alpha  \right)\,,\\
    {\texttt j}_s^\alpha = \int \frac{d\vartheta}{2\pi} P(\vartheta)  n^\alpha(\vartheta) \left( \Tilde{h}^\alpha_s(\vartheta) - \alpha \Tilde{E}^\alpha(\vartheta) {\texttt q}_s^\alpha +\alpha \Tilde{P}^\alpha(\vartheta){\texttt j}_s^\alpha  \right)\,. 
\end{align}
Introducing the tilded charges and currents $\tilde{\texttt q}_s^\alpha, \tilde{\texttt \j}_s^\alpha $ defined in an obvious way from the integration of the corresponding tilded eigenvalues, we obtain again a system of two equations: 
\begin{equation}
\label{eq:expressionnoice}
    {\texttt j}_s^\alpha = \tilde{\texttt \j}_s^\alpha -\alpha {\texttt q}_s^\alpha \tilde{\texttt \j}_1^\alpha + \alpha {\texttt j}_s^\alpha \tilde{\texttt \j}_{-1}^\alpha \qquad \mathrm{and}\qquad 
         {\texttt q}_s^\alpha = \tilde{\texttt q}_s^\alpha -\alpha {\texttt q}_s^\alpha \tilde{\texttt q}_1^\alpha + \alpha {\texttt j}_s^\alpha \tilde{\texttt q}_{-1}^\alpha\,.
\end{equation}
This system is very similar to (\ref{34}), to which indeed it specialises when the tilde operation is trivial, namely when the unperturbed theory is free. These equations can be solved easily to give the final expressions:
\beq 
\begin{split}
       \qquad  {\texttt q}_s^\alpha &=\frac{\Tilde{ {\texttt q}}_s^\alpha + \alpha \tilde{ {\texttt q}}_{-1}^\alpha \Tilde{ {\texttt \j}}_s^\alpha- \alpha \tilde{{\texttt \j}}_{-1}^\alpha \Tilde{ {\texttt q}}_s^\alpha}
         {1+ \alpha (\Tilde{ {\texttt q}}_1^\alpha -\Tilde{{\texttt \j}}_{-1}^\alpha )+ \alpha^2 (\Tilde{{\texttt \j}}_1^\alpha \Tilde{ {\texttt q}}_{-1}^\alpha-\Tilde{{\texttt \j}}_{-1}^\alpha \Tilde{ {\texttt q}}_1^\alpha)}\,, \qquad  \\
        \qquad {\texttt j}_s^\alpha &= \frac{\Tilde{{\texttt \j}}_s^\alpha + \alpha \tilde{ {\texttt q}}_1^\alpha \Tilde{{\texttt \j}}_s^\alpha- \alpha \tilde{\texttt \j}_1^\alpha \Tilde{ {\texttt q}}_s^\alpha} {1+ \alpha (\Tilde{ {\texttt q}}_1^\alpha -\Tilde{{\texttt \j}}_{-1}^\alpha )+ \alpha^2 (\Tilde{{\texttt \j}}_1^\alpha \Tilde{ {\texttt q}}_{-1}^\alpha-\Tilde{{\texttt \j}}_{-1}^\alpha \Tilde{ {\texttt q}}_1^\alpha)}\,.\qquad
    \label{eq:TildeSystem}
    \end{split}
\eeq
Any situation in which the tilded quantities can be explicitly expressed in terms of the unperturbed ones leads to an exact solution, as for the free fermions seen earlier. However, in most cases, the unperturbed quantities are only accessible numerically. There are however two cases where simplifications occur, namely the equilibrium situation (either the free massive case  or the massless general case) and the partitioning protocol. 

\subsection{Equilibrium}
As we have seen for the free fermion earlier, at equilibrium the TBA and dressing equations are identical to those in an unperturbed theory at inverse temperature $\hat{\beta}$. Therefore the currents and densities calculated using $n^\alpha(\vartheta) $ can be exactly calculated from $\hat{n}^0(\vartheta)$. It is then clear that the operation \eqref{eq:tilde} is exactly equal to the dressing operation in the unperturbed theory at the modified temperature. This means that we can identify $\tilde{\texttt q}_s^\alpha=\hat{\texttt q}_s^0$ and $\tilde{\texttt \j}_s^\alpha=\hat{\texttt \j}_s^0$. 

If we just consider the even charges, associated with one-particle eigenvalues that are even functions of $\vartheta$, then all currents ${\texttt j}_s^\alpha = 0$, and from \eqref{eq:TildeSystem} we obtain again
(\ref{freef}). We find that this formula holds for interacting models at any temperature, and it is exact. 

A situation where we can make more progress analytically is at critical points. Below, we consider again the massless limit of the free fermion theory. 
Even though this section is about interacting theories, the free fermion provides a useful benchmark for equation (\ref{freef}), since in the free fermion case the averages of currents and densities in the massless limit are analytically accessible and can then be compared to (\ref{freef}). Furthermore, free fermion averages in the massless limit display the same universal dependence on $\hat{\beta}$ that is found for more general CFTs, albeit with a different numerical prefactor \cite{MazzoniUnpublished}. 

We observe that the dressing operation is simply $h_s^{\rm{dr},\alpha}(\vartheta)=h_s(\vartheta)-\alpha E(\vartheta) {\texttt q}^\alpha_s$. The charges can be easily computed:
\beqa 
{\texttt q}_s^\alpha &=& \frac{m}{2\pi} \int_{-\infty}^\infty \cosh\vartheta' n^\alpha(\vartheta') h_s^{\rm dr, \alpha}(\vartheta')d\vartheta' \\
&=& \frac{m}{2\pi} \int_{-\infty}^\infty \cosh\vartheta' n^\alpha(\vartheta') h_s(\vartheta')d\vartheta'-\alpha {\texttt q}^\alpha_s \frac{m^2}{2\pi} \int_{-\infty}^\infty \cosh^2\vartheta' n^\alpha(\vartheta')d\vartheta'\nonumber\\
&=&  \frac{m^{s+1}}{2\pi} \int_{-\infty}^\infty \frac{\cosh\vartheta'\cosh(s\vartheta')}{1+e^{m\hat{\beta}\cosh\vartheta'}}d\vartheta' -\alpha {\texttt q}^\alpha_s \frac{m^2}{2\pi} \int_{-\infty}^\infty \frac{\cosh^2\vartheta'}{1+e^{m\hat{\beta} \cosh\vartheta'}}d\vartheta'  \,.
\eeqa 
Here we have taken the even charge eigenvalue $h_s(\vartheta)=m^s\cosh(s\vartheta)$, but we could have taken a combination of $\cosh(s\vartheta)$ and $\sinh(s\vartheta)$ functions without changing the essence of the calculation. For $m\hat{\beta} > 0$, the denominator in the integrals admit the geometric series expansion
\beqa 
\frac{1}{1+e^{m\hat{\beta}\cosh\vartheta'}}=\sum_{n=1}^\infty (-1)^{n+1} e^{-n m\hat{\beta}\cosh\vartheta'}\,,
\label{49}
\eeqa 
and thus we can again make use modified Bessel functions, this time of higher order, to rewrite the integrals:
\beqa
\int_0^\infty \cosh \vartheta \cosh(s\vartheta) e^{-A \cosh\vartheta} d\vartheta&=& \int_0^\infty \cosh (s+1)\vartheta \, e^{-A \cosh\vartheta} d\vartheta \nonumber \\
&-&\int_0^\infty \sinh \vartheta \sinh(s\vartheta) e^{-A \cosh\vartheta} d\vartheta\nonumber\\
&=& K_{s+1}(A)-\frac{s}{A} K_s(A)\,,\quad \mathrm{for}\quad A\neq 0\,.
\label{50}
\eeqa 
Using now the asymptotic expansion $K_s(x) \sim \frac{\Gamma(s) 2^{s-1}}{x^s}$ for $x \sim 0$, the exact expression for ${\texttt q}_s^\alpha$ as sum of modified Bessel function can be rewritten in terms of the Riemann zeta function in the $m \ll 1$ limit:
\beqa 
\label{casinofolle}
{\texttt q}_s^\alpha&=&\frac{m^{s+1}}{\pi} \sum_{n=1}^\infty (-1)^{n+1} \left(K_{s+1}(n m\hat{\beta})-\frac{s}{n m\hat{\beta}} K_s(n m\hat{\beta})\right)\nonumber\\
&& -\frac{m^2 \alpha {\texttt q}_s^\alpha}{\pi} \sum_{n=1}^\infty (-1)^{n+1} \left(K_{2}(n m\hat{\beta})-\frac{1}{n m\hat{\beta}} K_1(n m\hat{\beta})\right)\nonumber \label{ciao}\\
&\approx & \frac{m^{s+1}}{\pi} \sum_{n=1}^\infty (-1)^{n+1} \left(\frac{s! 2^s}{(n m\hat{\beta})^{s+1}} -\frac{s!2^{s-1}}{(n m\hat{\beta})^{s+1}} \right) -\frac{\alpha m^2 {\texttt q}_s^\alpha}{\pi} \sum_{n=1}^\infty (-1)^{n+1} \left(\frac{ 2}{(n m\hat{\beta})^{2}} -\frac{1}{(n m\hat{\beta})^{2}} \right)\nonumber\\
&=& \frac{1}{\pi} \sum_{n=1}^\infty (-1)^{n+1} \left(\frac{s! 2^{s-1}}{(n\hat{\beta})^{s+1}} - \frac{\alpha q_s^\alpha}{(n\hat{\beta})^{2}} \right)\qquad \mathrm{for}\qquad m\ll 1\,.
\eeqa 
As expected, the mass dependence cancels out in the massless limit.
Using the known sum:
\begin{equation}
    \sum_{n=1}^{\infty}\frac{(-1)^{n+1}}{n^{s+1}}=\zeta(s+1)(1-2^{-s}),
\end{equation}
we obtain the final expression:
\beq 
{\texttt q}_s^\alpha=\frac{s! 2^{s-1}\zeta(s+1) (1-2^{-s})}{\pi\hat{\beta}^{s+1}} -\frac{\alpha {\texttt q}_s^\alpha  \pi}{12 \hat{\beta}^{2}}\,,
\eeq 
from which finally we can read
\beq 
{\texttt q}_s^\alpha=\frac{s! 2^{s-1}\zeta(s+1) (1-2^{-s})}{\pi\hat{\beta}^{s+1}\left(1+\frac{\alpha  \pi}{12\hat{\beta}^{2}}\right)} \,.
\label{27}
\eeq 
This expression can be compared to (\ref{freef}) using the results found in \cite{Mazzoni,MazzoniUnpublished} for the free fermion density averages:
\begin{equation}
    {\texttt q}_s^0 = \frac{s! 2^{s-1}}{ \pi \beta^{s+1}}(1-2^{-s}) \zeta(s+1)\,,
    \label{comparisonfreetheory}
\end{equation}
and in particular ${\texttt q}^0_1 = \frac{\pi}{12\beta^2} = {\texttt j}^0_{-1}$. Substituting these free results (evaluated at $\hat{\beta}$ ) into \eqref{freef}, we obtain (\ref{27}) as expected.
Although these results are only valid in the conformal limit, expression \eqref{freef} is valid for all values of $m$. Therefore, it provides a starting point for obtaining perturbative results beyond the CFT point. 

\subsection{Partitioning Protocol}
\label{42}
 Out-of equilibrium configurations are harder to treat because the effect of the perturbation can no longer be absorbed into redefinition of temperature. The occupation function depends on $\beta$, $\vartheta_\star^\alpha$ and $\alpha$ as shown in equations (\ref{n_partitioning}) and (\ref{veff1}). The main difference with the equilibrium case is that $\vartheta^\alpha_\star \neq \vartheta_\star^0$, and therefore in general $n^\alpha(\vartheta)\neq \hat{n}^0(\vartheta)$ and the relationship between perturbed and unperturbed quantities is not obvious a priori.
There are, however, some approximations that can be made close to a critical point, a fact that will allow us to once again rely on formulae we have obtained previously. 

\begin{figure}[h!]
    \centering
    \includegraphics[width=\textwidth]{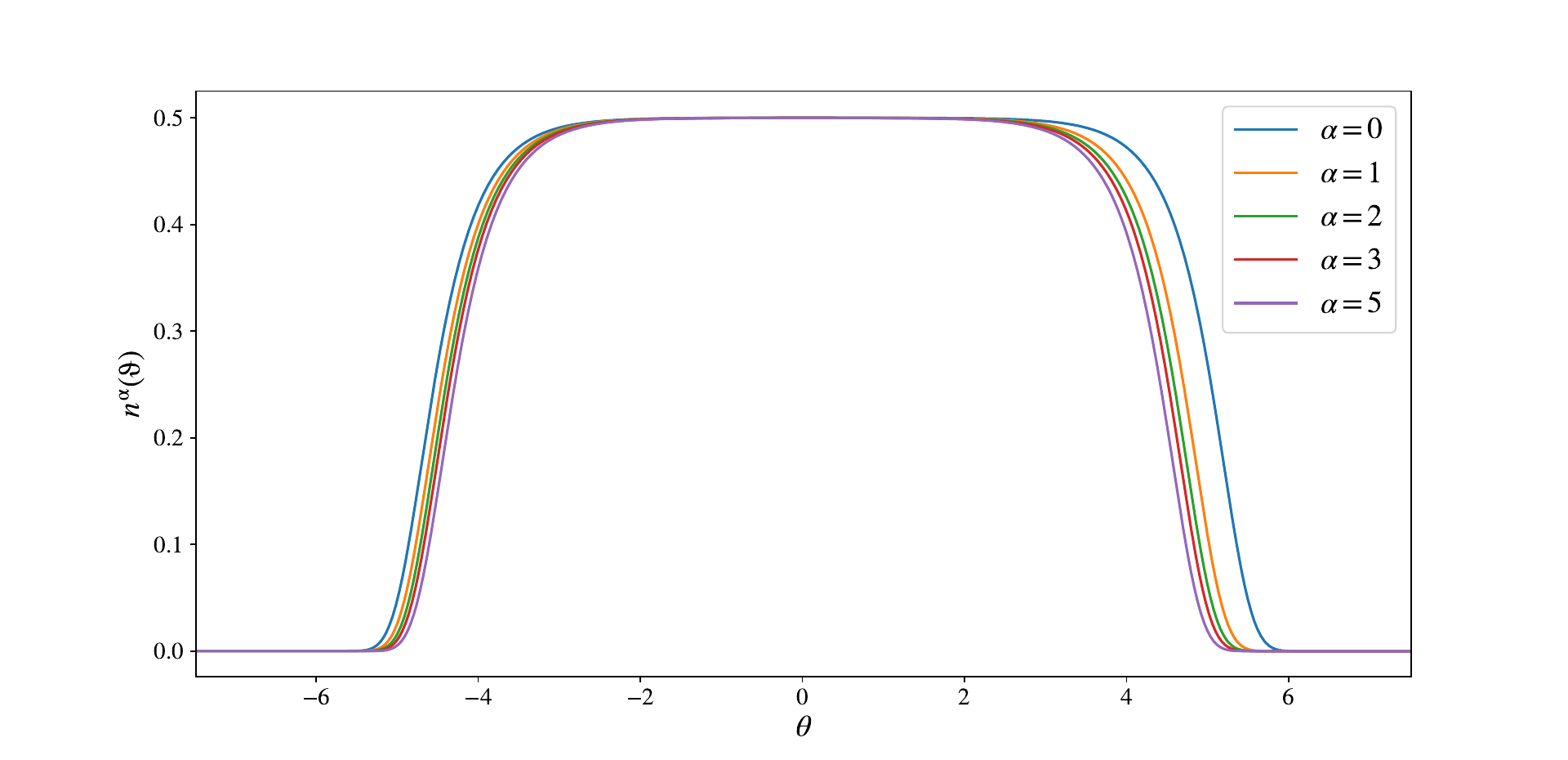}
    \caption{Plateau structure arising in the perturbed Lee-Yang scaling model for $\beta_R/\beta_L=3$ and $m=0.0001$, for different values of $\alpha$. For $m\ll 1$ the theory is near-critical and we recover $n^\alpha(\theta)=\hat{n}^0(\theta)$. There is the same type of plateau structure as for $\alpha=0$ up to a redefinition of right/left temperatures. Note that $n^\alpha(\vartheta)\neq n^\alpha(-\vartheta)$, although this effect is less evident for smaller $m$. The height of the plateau is $\log\Phi$, where $\Phi=\frac{1+\sqrt{5}}{2}$ is the golden ratio, as predicted by the constant TBA equations \cite{Zamolodchikov:1989cf}.}
    \label{fig1}
\end{figure}

From the definition of the effective velocity (\ref{veff1}), we see that the value(s) of $\vartheta_\star^\alpha$ correspond to the solutions of
\begin{equation}
P^{\rm dr,\alpha}(\vartheta_\star^\alpha) = \tilde{P}_\alpha(\vartheta_\star^\alpha) -\alpha {\texttt q}_{-1}^\alpha \Tilde{E}^\alpha(\vartheta_\star^\alpha) +\alpha {\texttt j}_{-1}^\alpha \Tilde{P}^\alpha(\vartheta_\star^\alpha), = 0,
\end{equation}
where we used (\ref{eq:TildeDressingEquation}). 
This gives:
\begin{equation}
    \frac{\tilde{P}^\alpha(\vartheta_\star^\alpha)}{\Tilde{E}^\alpha(\vartheta_\star^\alpha)} = \frac{\alpha {\texttt q}_{-1}^\alpha }{1+ \alpha {\texttt j}_{-1}^\alpha }.
\end{equation}
In general, the solution of this equation will lead to a value of $\vartheta_\star^\alpha$ different from that of the unperturbed case. However, we can argue that  things simplify at and near a critical point.  This is due to the form of the occupation functions, which in the massless limit (that is the limit in which the theory becomes a $T\Bar{T}$ deformed CFT) develop a large plateau centered around $\vartheta=0$. In the partitioning protocol, $n^0(\vartheta)$ is generally not symmetric with respect to $\vartheta$, but it becomes very nearly so when both $\beta_R, \beta_L$ are large. Once the temperatures are high enough for $n_L^0(\vartheta), n_R^0(\vartheta)$ to develop the asymptotic high temperature plateau, the value of the connection point $\vartheta_\star^0$ ceases to matter (as long as it falls within the plateau region). This intuitive idea extends to the perturbed case too and can be tested numerically as we see in Fig.~\ref{fig1}. Here, we reach the UV limit by tuning the mass scale instead (which is more natural in the presence of the two scales $\beta, \alpha$).
In the perturbed case, we observe that the larger $\alpha$ is, the smaller we need to make $m$ in order to see a well-developed plateau. 

In summary, at the critical point the same formulae (\ref{qs})-(\ref{js}) hold for any interacting theory with a single particle spectrum. This is one of the main results of this work and generalises the findings of \cite{MedenjakLong,MedenjakShort} to higher spin charges and currents. The extension to many-particle theories is straightforward as long as the scattering is diagonal, as discussed in Appendix \ref{appA}, while non-diagonal TBA systems require in general a case by case analysis.


The rationale behind our approach is to be as general as possible and to look at the conformal limit only when it is impossible to obtain analytical results otherwise. In this way we can make additional observations, namely that the results \eqref{qs}-\eqref{js} are valid also when $m \gg \alpha$ or when $\beta \gg \alpha $ in out of equilibrium configurations, and for any value of the mass at equilibrium. A derivation of equations \eqref{qs}-\eqref{js} starting directly from the CFT, i.e. the approach adopted in \cite{MedenjakLong,MedenjakShort}, is presented in Appendix \ref{appC}. 

\subsection{General CFT}
\label{subCLR}
A benchmark for some of our results is the work \cite{MedenjakLong}, where the energy density and current of a generic $\TTb$-perturbed CFT were computed in the NESS arising after a partitioning protocol. In our case, we just need to take the formulae \eqref{qs}-\eqref{js} for $s=1$ and substitute in the values for unperturbed CFT \cite{bernard2012energy, Bernard:2013aru, bernard2016conformal}, which are given by
\beq
{\texttt j}^0_1 = {\texttt q}^0_{-1} = \frac{c \pi }{12}\left({T_L^2}-{T_R^2}\right) \qquad \mathrm{and}\qquad
{\texttt q}^0_1 = {\texttt j}^0_{-1} = \frac{c \pi }{12}\left({T_L^2}+T_R^2\right)\,.
\eeq
The contributions depending on $\beta_R$ and $\beta_L$ separately, a property that is also found for higher currents and densities \cite{Mazzoni,MazzoniUnpublished} when $\alpha=0$. In the presence of a  $\TTb$ perturbation, this strict separation no longer holds. Substituting into \eqref{qs}-\eqref{js} for $s=1$  we obtain:
\begin{eqnarray}
    {\texttt j}_1^\alpha = \frac{{\texttt j}_1^0}{1-\alpha^2 ({\texttt q}_1^0)^2+\alpha^2 ({\texttt j}_1^0)^2}={\frac{c \pi }{12}}c_{LR}\left({\hat{T}_L^2}-{\hat{T}_R^2}\right)\,,
    \label{eq:jemedenjak}
\end{eqnarray}
and similarly, 
\begin{eqnarray}
    {\texttt q}_1^\alpha = \frac{{\texttt q}_1^0 +\alpha \left(({\texttt j}_1^0)^2 - ({\texttt q}_1^0)^2\right)}{1-\alpha^2 ({\texttt q}_1^0)^2+\alpha^2 ({\texttt j}_1^0)^2} = {\frac{c \pi }{12}}c_{LR}\left({\hat{T}_L^2}+ {\hat{T}_R^2} - \frac{\pi c \alpha}{3} \hat{T}_R^2 \hat{T}_L^2 \right)\,,
    \label{eq:qemedenjak}
\end{eqnarray}
with 
\beq 
c_{LR}:= \frac{1}{1-(\frac{\alpha \pi c}{6})^2\hat{T}_R^2 \hat{T}_L^2}\,.
\eeq 
These results agree with \cite{MedenjakLong} where a different approach was employed (that of massless TBA), and therefore this provides a substantial confirmation of the validity of our formulae in the CFT limit. Compared to the unperturbed case and even after accounting for the redefinition of the temperatures, the factor $c_{LR}$ introduces a mixing of right and left variables.
 
Let us now consider higher spin quantities. In \cite{Mazzoni,MazzoniUnpublished} it was shown that the NESS averages in the CFT limit are:
\beqa
\label{eq:mazzonicharges}
    {\texttt j}_s^0 = G(s) \left(T_L^{s+1} -T_R^{s+1}\right)\qquad \mathrm{and}\qquad
   {\texttt q}_s^0 =G(s) \left(T_L^{s+1} +T_R^{s+1}\right)\,.
\eeqa
The proportionality constant\footnote{In \cite{MazzoniUnpublished} we adopted a different normalisation of the constant, so that the coefficient $\mathcal{C}(s)$ defined therein is related to $G(s)$ by $G(s)=\frac{s\,2^s\,\pi}{24}\mathcal{C}(s)$.} $G(s)$ is a theory-specific quantity that can be computed from the TBA equations and for which no general closed expression is known, except for the free fermion case and for $s=1$ where $G(1)=\frac{c\pi}{12}$. Substituting these formulae into \eqref{qs}-\eqref{js} we obtain:
\beqa
     {\texttt j}_s^\alpha &=& G(s)c_{LR}\left(\left(\hat{T}_L^{s+1} -\hat{T}_R^{s+1}\right) + \frac{\alpha \pi c}{6} (\hat{T}_L^{s+1} \hat{T}_R^2 -\hat{T}_R^{s+1}\hat{T}_L^2) \right)\,,
     \label{eq:finalcurrent}\\
      {\texttt q}_s^\alpha &=& G(s)c_{LR}\left(\left(\hat{T}_L^{s+1} +\hat{T}_R^{s+1}\right) - \frac{\alpha \pi c}{6}(\hat{T}_L^{s+1} \hat{T}_R^2 +\hat{T}_R^{s+1}\hat{T}_L^2) \right)\,.
      \label{eq:finalcharge}
\eeqa
As anticipated in the Introduction, these two formulae are the main finding of  our work. The results show that the presence of the $\TTb$ deformation breaks the left-right separation also for the higher currents and densities.  Note that these are the currents associated to even charges with one-particle eigenvalue $ m^s \cosh( s\vartheta)$. Similar formulae can be written for the odd charges, namely
\beqa 
        {\texttt q}_{-s}^\alpha &=& G(s)c_{LR}\left(\left(\hat{T}_L^{s+1} -\hat{T}_R^{s+1}\right) - \frac{\alpha \pi c}{6}(\hat{T}_L^{s+1} \hat{T}_R^2 -\hat{T}_R^{s+1}\hat{T}_L^2) \right)\,, \\
         {\texttt j}_{-s}^\alpha &=& G(s)c_{LR}\left(\left(\hat{T}_L^{s+1} +\hat{T}_R^{s+1}\right) + \frac{\alpha \pi c}{6} (\hat{T}_L^{s+1} \hat{T}_R^2 +\hat{T}_R^{s+1}\hat{T}_L^2) \right)\,. 
    \eeqa 
From this we obtain the value of the momentum current which was also found in \cite{MedenjakLong}. As a concluding remark, we observe that the symmetry relations ${\texttt q}_s={\texttt j}_{-s}$ and ${\texttt j}_{s}={\texttt q}_{-s}$, which hold in unperturbed CFT, are violated here, again because of the interaction between right and left movers introduced by the perturbation. The only exception is ${\texttt j}^\alpha_1={\texttt q}^\alpha_{-1}$: however, while in the unperturbed theory this equation holds for massive theories as well, when $\alpha\neq 0$ it is true only in the conformal limit.
\section{Scaling Function}
\label{scaling}
We have already defined the function $c^\alpha$ in (\ref{eq:c_function_medenjak}), that is the counterpart of the UV central charge in the $\TTb$ perturbed theory. In the presence of a deformation, $c^\alpha$  is no longer a constant, but it is a function of $\beta^2/\alpha$. In the unperturbed theory one defines the TBA scaling function, which away from the critical point is a function of $r:=m\beta$, through $E_0:=-\frac{\pi c^0(r)}{6\beta}$. At equilibrium, we know that all TBA quantities in the $\TTb$-perturbed theory are identical to those in the unperturbed theory evaluated at a modified inverse temperature $\hat{\beta}$. Thus, we can introduce a new scale $\hat{r}=m\hat{\beta}$ and express the free energy using our standard notation as $E_0^\alpha=\hat{E}_0$, or equivalently:
\beq 
\hat{\beta} \,c^\alpha(r,r')=\beta \, c^0(\hat{r})\,.
\label{66}
\eeq 
This gives a relationship between the two scaling functions.  We have written $c^\alpha(r,r')$ to emphasise that the scaling function of the perturbed theory depends on two independent dimensionless scales, with $r'=m^2\alpha$. If we then employ the relation {\iffalse\color{red}\fi $\hat{\beta}=\beta-\alpha E_0^\alpha = \beta+ \frac{\pi \alpha c^0(\hat{r})}{6\hat{\beta}}$}, we can further eliminate any explicit dependence on $\beta$ and write
\begin{equation}
\label{eq:calpha}
    c^\alpha(r,r')=c^0(\hat{r})\left(1-\frac{\alpha \pi}{6}\frac{c^0(\hat{r})}{\hat{\beta}^2}\right)\,.
\end{equation}
We can now try to say something about the asymptotic and monotonicity properties of $c^\alpha(r,r')$ from those of $c^0(\hat{r})$. 
The main properties of the latter are the same as for Zamolodchikov's $c$-function, as demonstrated by the $c$-theorem \cite{Zamolodchikov:1986gt}: it is a monotonic function of $\hat{r}$ with $\frac{\partial c^0(\hat{r})}{\partial \hat r} \leq 0$ and asymptotic values $c^0(0)=c$ (UV limit) and $c^0(\infty)=0$ (IR limit for a massive IQFT). 

An interesting property is that $c^\alpha(r,r') \leq  c^0(r)$ for every choice of the parameters. This follows immediately when noting that $\hat{\beta}>\beta$ (for our choice $\alpha>0$) and that $c^0(r)$ is a decreasing function of $r$. The properties of the $c$-function, and more generally RG flows in $\TTb$-perturbed theories, have been investigated in detail in \cite{LeClair:2021opx,LeClair:2021wfd,LeClair_2022,Ahn:2022pia}. It has also been recently shown that functions of $r'$ that flow monotonically from the value $c$ to zero can be defined employing form factors of the stress-energy tensor and the branch point twist field in $\TTb$-deformed theories \cite{longpaper,PRL,ourentropy}.

\subsection{Limiting Values: Large and Small Mass Limits}
The first property we can infer and which we already discussed earlier, is that for $\hat{r}=0$ we obtain a value of $c^\alpha(r,r')$ which is no longer a constant, but a function of $\alpha/\hat{\beta}^2$ (see Eq. (\ref{eq:c_function_medenjak}) which is a rewriting of (\ref{eq:calpha}) in terms of $\beta$ for this particular limit). One way to make sense of this with the present formula is to take the limit $\hat{r}\rightarrow 0$ by taking $m\rightarrow 0$ while keeping both $\beta$ and $\alpha$ finite and fixed. Concerning the limit $\hat{r}\rightarrow \infty$, we can again carry out this by sending $m\rightarrow \infty$ while keeping $\alpha, \beta$ fixed and finite. In this case the formula (\ref{eq:calpha}) immediately gives $c^\alpha(\infty,\infty)=0$. 

Thus, while the small mass limit yields a universal result which  depends on the scale $\alpha/\hat{\beta}^2$ \cite{Cavaglia:2016oda}, the large mass limit is the same as for the unperturbed theory. These two behaviours are consistent with other findings about $\TTb$ perturbations. 
A simple argument is that the effect of the $\TTb$ perturbation is to cause the UV theory to be ill-defined, or at least to be no longer a local QFT, so it is a short-distance effect. In the language of generalised hydrodynamics, we can also think of the perturbation as assigning finite length to elementary degrees of freedom \cite{Doyon:2021tzy,Cardy:2020olv,Doyon:2023bvo}. What emerges from these interpretations is that $\TTb$ perturbation should play an important role at short distances (or, alternatively, small mass) whereas in the infrared (for large distances/mass) the effect of the $\TTb$ perturbation is not seen. This is consistent with the asymptotic properties of the $c$-function and of correlations functions \cite{longpaper}.

It is also clear from (\ref{eq:calpha}) that the properties of $c^\alpha(r,r')$ when either $r=0$ or $r'=0$ (but not both), are non-trivial. For instance, if $r=0$ but $r'\neq 0$ then $\beta=0$ and $\hat{\beta}=-\alpha \hat{E}_0=\frac{\pi c^0(- m \alpha \hat{E}_0)} {-6 \hat{E}_0 }$ which gives a recursive relation involving the ground state energy
\beq 
6\alpha \hat{E}_0^2 = c^0 (-m \alpha \hat{E}_0)\,.
\eeq 
This relation can be exploited for instance in the few special cases where the function $\hat{E}_0$ is known explicitly and in perturbative calculations in $m$ or $\alpha$ (see also Appendix \ref{appD}).

\subsection{Monotonicity Properties}
Given the discussion above, we expect that the function $c^\alpha(r,r')$ should also be a monotonic function, albeit not with respect to the variable $r$ but with respect to the mass scale, which is the scale that allows the theory to flow from the CFT fixed point of the original unperturbed model to the infrared. This can be shown starting from equation (\ref{66}), namely
\beqa
\frac{\partial c^\alpha(r,r') }{\partial m } &=& \frac{\beta}{\hat{\beta}^2} \left(\frac{\partial c^0(\hat{r})}{\partial m } \hat{\beta} - c^0(\hat{r}) \frac{\partial \hat{\beta}}{\partial m} \right)
= \beta \frac{\partial c^0(\hat{r})}{\partial \hat{r} } + \frac{\beta}{\hat{\beta}^2} \frac{\partial\hat{\beta}}{\partial m } \left(\frac{\partial}{\partial \hat{r}} \left(\frac{c_0(\hat{r})}{\hat{r}}\right)\hat{r}^2\right)\,.
\label{eq:intermediate}
\eeqa
The mass derivative of the modified temperature can be simply obtained from the definition (\ref{eq:defBeta}):
\beqa
    \frac{\partial\hat{\beta}}{\partial m}=-\alpha \frac{\partial E_0^\alpha}{\partial m} = \frac{\alpha \pi}{6\beta}\frac{\partial c^\alpha(r,r')}{\partial m},
\eeqa
which when substituted back in \eqref{eq:intermediate} yields:
\begin{equation}
\label{monotonicityofscalingfunction}
    m \frac{\partial c^\alpha(r,r') }{\partial m } = \left(\frac{r}{1-\frac{\pi r'}{6} \frac{\partial}{\partial \hat{r}}\left(\frac{c^0(\hat{r})}{\hat{r}}\right)}\right) \frac{\partial c^0(\hat{r})}{\partial \hat{r}}\,.
\end{equation}
Since $\frac{\partial}{\partial \hat{r}}\left(\frac{c_0(\hat{r})}{\hat{r}}\right) <0$ the term in brackets is positive, and this proves that $c^\alpha(r,r')$ and $c^0(\hat{r})$ have the same monotonicity. Note however the explicit dependence of (\ref{monotonicityofscalingfunction}) on $r'$. 
The flow along the direction identified by the mass is the natural generalisation of the standard RG flow in the unperturbed case, with the difference that in the present situation we have both relevant and irrelevant deformations, and a variation of $m$ produces a flow along the relevant direction with the irrelevant deformation being \lq\lq carried along" in the process.

We may study the monotonicity properties with respect to different flows. For instance, one can show that $c^\alpha(r,r')$ is also monotonically decreasing as a function of $\alpha$. The calculation is analogous to (\ref{eq:intermediate}) and yields a very similar result, namely 
\beqa
\frac{\partial c^\alpha(r,r') }{\partial \alpha } =\frac{\beta}{\hat{\beta}^2} \left(\frac{\partial c^0(\hat{r})}{\partial \alpha } \hat{\beta} - c^0(\hat{r}) \frac{\partial \hat{\beta}}{\partial \alpha} \right)=
m r \frac{\partial \hat{\beta}}{\partial \alpha} \frac{\partial}{\partial \hat{r}} \left(\frac{c^0(\hat{r})}{\hat{r}}\right)\,.
\label{eq:alpha}
\eeqa
Since 
\beq 
\frac{\partial\hat{\beta}}{\partial \alpha }=-E_0^\alpha-\alpha \frac{\partial E_0^\alpha}{\partial \alpha} =\frac{\pi c^\alpha(r,r')}{6\beta}+ \frac{\alpha \pi}{6\beta}\frac{\partial c^\alpha(r,r')}{\partial \alpha}\,, 
\eeq 
we have 
\beq 
\frac{1}{m^{2}}\frac{\partial c^\alpha(r,r')}{\partial \alpha}= \frac{\pi}{6}\left(\frac{c^{\alpha}(r,r')}{1-\frac{\pi r'}{6} \frac{\partial}{\partial \hat{r}}\left(\frac{c^0(\hat{r})}{\hat{r}}\right)}\right) \frac{\partial}{\partial \hat{r}}\left(\frac{c^0(\hat{r})}{\hat{r}}\right) < 0\,,
\eeq
where the inequality follows from the fact that the function in brackets is positive. 
In contrast, there is no monotonicity with respect to $\beta$:
\begin{equation}
   \frac{1}{m} \frac{\partial c^\alpha(r,r')}{\partial \beta} = \frac{c^0(\hat{r})}{\hat{r}} +  r \frac{\frac{\partial}{\partial \hat{r}}\left(\frac{c^0(\hat{r})}{\hat{r}}\right)}{1-\frac{\pi r'}{6}\frac{\partial}{\partial \hat{r}}\left(\frac{c_0(\hat{r})}{\hat{r}}\right) }\,,
    \label{exp}
\end{equation}
since the first term on the r.h.s. is always positive and the second term is always negative. 
A simple study of the asymptotics indeed shows that the derivative does change sign: for $\beta \approx 0$ the first term dominates, since $\hat{\beta}$ is finite as $\beta \to 0$ and the second term  is very small ($r \approx 0$). On the other hand, if $\beta \to \infty$ then $\hat{\beta}\approx \beta$, in which case the first term in (\ref{exp}) tends to zero while the second term grows linearly with $\beta$. Alternatively, by approximating the denominator of the second term we can write:
\begin{equation}
    \frac{1}{m} \frac{\partial c^\alpha(r,r')}{\partial \beta} \approx \frac{c^0(r)}{r} + \frac{1}{r}\left(\frac{\partial c^0(r)}{\partial r} r - c^0(r)\right) =  \frac{\partial c^0(r)}{\partial r} \leq 0 \qquad \mathrm{for}\qquad \beta \gg 1\,.
     \label{mono}
\end{equation}
The non-monotonicity of $c^\alpha(r,r')$ with respect to $\beta$ further emphasises the fact that the parameter $r$ is no longer the only dimensionless scale in the problem. Note that the result (\ref{mono}) is exact for any $\beta$ if $r'=0$, as monotonicity is restored when there is no perturbation. 


\section{Numerical Results}\label{numerics}
\begin{figure}
\centering
\begin{subfigure}[b]{\textwidth}
   \includegraphics[width=1\linewidth]{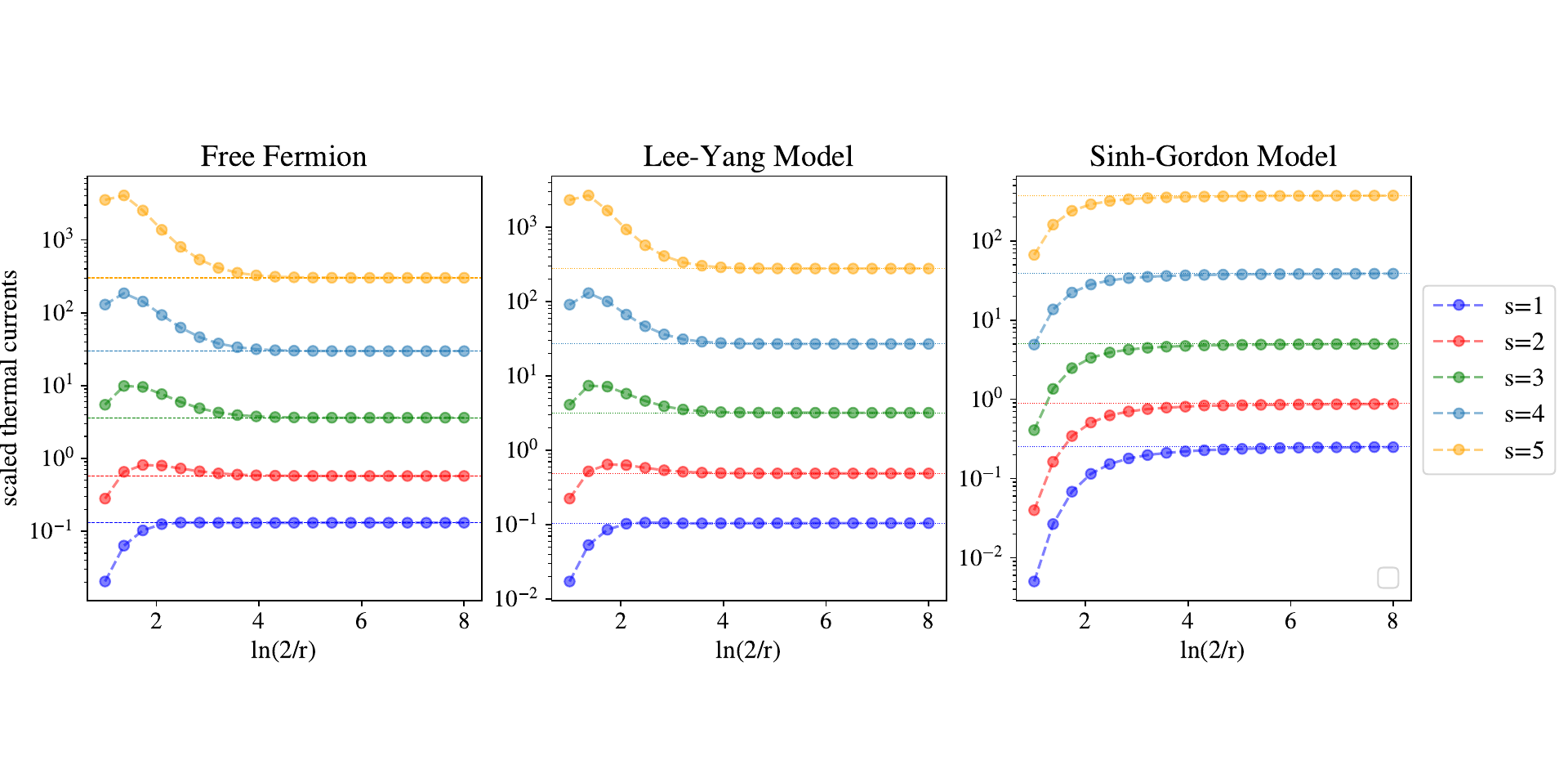}
   \caption{Currents for $\alpha=1$, $\beta_L = 1/3$, $\beta_R=1$}
   \label{fig:currents1} 
\end{subfigure}
\begin{subfigure}[b]{1\textwidth}
   \includegraphics[width=1\linewidth]{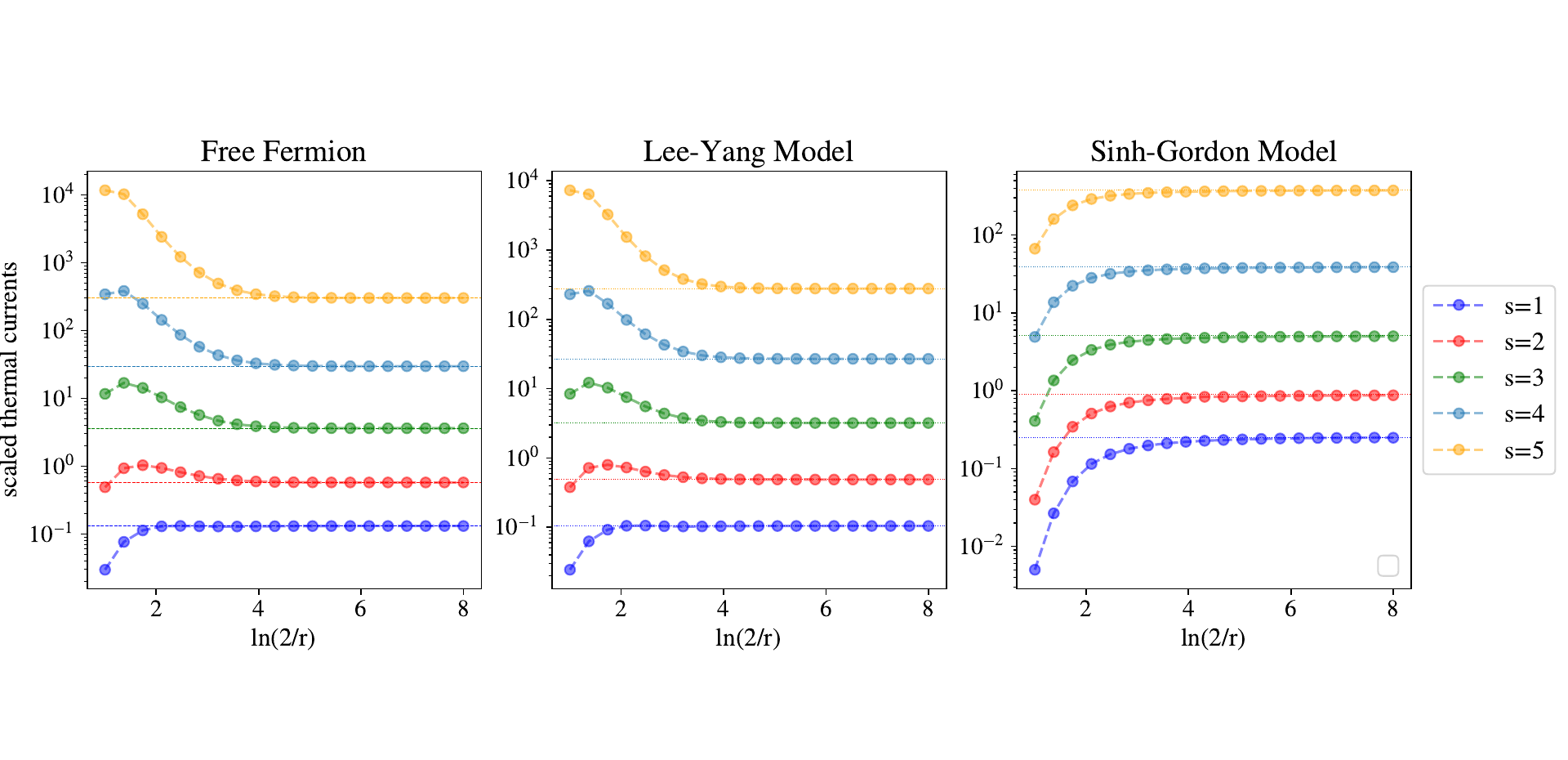}
   \caption{Currents for $\alpha=3$, $\beta_L = 1/2$, $\beta_R=1$}
   \label{fig:currents2}
\end{subfigure}
\caption{Normalised currents \ref{jnorm} for different values of $\alpha$ and of the temperatures. The curves are obtained by varying the mass: the massless limit is attained at large $\ln{(2/r)}$. The horizontal dashed lines are the values G(s), which are evaluated numerically as done in \cite{MazzoniUnpublished}. }
\end{figure}

In this section we perform numerical tests to check the overall validity of our theoretical predictions, in particular of the two expressions \eqref{eq:finalcurrent} and \eqref{eq:finalcharge}. These expressions predict the behaviour of densities and currents in the $m \to 0$ limit, which can be tested in a rather straightforward way. First of all, one can simulate a partitioning protocol by solving iteratively the TBA equations (see for instance \cite{MazzoniUnpublished}) to obtain the numerical values of the currents $j_\text{simul}$ for different values of $m$. Since we are interested in the NESS currents, we evaluate them for $\xi=0$. The currents can then be normalised in order to obtain a quantity which in the $m \to 0$ limit do not depend on the choice of $\alpha$ nor on the right and left effective temperatures:
\begin{equation}
    j_\text{norm} = \frac{j_\text{simul}}{c_{LR}\left(\left(\hat{T}_L^{s+1} -\hat{T}_R^{s+1}\right) + \frac{\alpha \pi c}{6} (\hat{T}_L^{s+1} \hat{T}_R^2 -\hat{T}_R^{s+1}\hat{T}_L^2) \right) }\,.
    \label{jnorm}
\end{equation}
We focus on even currents without any loss of generality, {\iffalse\color{red}\fi since odd currents and even/odd charges are obtained numerically solving similar integral equations, and thus no particular difference or complication would arise}. In the conformal limit, we expect $j_\text{norm}$ to approach the value of $G(s)$. In order to verify that \eqref{eq:finalcurrent} and \eqref{eq:finalcharge} have the correct temperature dependence, it is enough to check that $j_\text{norm}$ is independent on $T_L$ and $T_R$ in the massless limit, since this means that through \eqref{jnorm} we are indeed removing all the temperature dependence. 
Numerical simulations are performed using the same three models which were considered in \cite{MazzoniUnpublished}, namely  the free fermion, the scaling Lee-Yang model and the sinh-Gordon model at the self-dual point. These are all single-particle IQFTs which in the massless limit are described by CFTs of (effective) central charges $c_\text{eff}=\frac{1}{2}$, $\frac{2}{5}$ and 1 respectively. The results of the numerical evaluations are shown in figures \ref{fig:currents1} and \ref{fig:currents2} for different values of the parameters. The plots show indeed that the asymptotic value which is reached is independent on the choice of the parameters and it is given by the value of $G(s)$, which is also evaluated numerically.

\section{Conclusions and Outlook}
\label{conclusion}
In this paper, we have studied the thermodynamic properties of massive IQFTs perturbed by $\TTb$ operator both at and away from equilibrium. Our main result are formulae for the averages of all local higher spin currents and densities. The averages in the perturbed theory are expressed as ratios of simple functions of the averages in the unperturbed theory.

In the conformal limit, equilibrium averages have been previously found \cite{Mazzoni,MazzoniUnpublished} and, as known since \cite{Smirnov:2016lqw,Cavaglia:2016oda},
the same formulae hold when the theory is $\TTb$-perturbed as long as the inverse temperature is replaced by a specific function of the latter and of the perturbation parameter. These formulae are here generalised to the out-of-equilibrium partitioning protocol. In this situation, the energy current and density were already known from \cite{MedenjakShort,MedenjakLong} and our results generalised the latter by extending them to generic integer values of spin. While in \cite{MedenjakShort,MedenjakLong} the authors considered the massless TBA equations as starting point, we start from the equations for a massive model and then take the massless limit of the currents and densities themselves. We obtain the universal dependence on the inverse temperatures for any CFT and find that, just as for the energy, for higher spins the $\TTb$ perturbation couples the right and left temperatures in a non-trivial fashion. There is, however, a spin-dependent proportionality coefficient that is theory-dependent and for the moment not accessible analytically, except for free fermions. 

For the massive free fermion several additional analytic computations are possible, as both the equilibrium and out-of-equilibrium averages can be obtained exactly. In addition, at equilibrium, it is possible to obtain a perturbative expansion of the effective inverse temperature $\hat{\beta}$  in terms of the original inverse temperature $\beta$. For small mass, this expansion can be resummed into a function which can be interpreted as a generalisation of Lambert's $W$-function.  Although the core of the paper focuses on theories with a single massive excitation, in the appendices we show the generalisation to many-particle spectra,  as well as the study of other thermodynamic/hydrodynamic quantities such as the effective velocities, which are known to have special properties in $\TTb$-perturbed models \cite{MedenjakLong,MedenjakShort} (e.g. superluminal propagation). The extension to more general $\TTb$-perturbations is also studied. 

The determination of the non-universal functions $G(s)$ in (\ref{critical}) remains one of the most interesting open problems. While $G(1)$ can be obtained exactly, for higher spins we have not yet found a closed formula which is valid for all CFT.
It has been shown \cite{Mazzoni,MazzoniUnpublished} that  $G(s)$ can be written in terms of integrals of TBA functions, of the same type as are encountered when relating the central charge to Roger's dilogarithm function. It appears natural that also these integrals might be solved by higher order polylogarithms. We hope to return to this problem in the future.

\medskip

 \noindent {\bf Acknowledgements:} 
 We thank Benjamin Doyon, D\'avid X. Horv\'ath, Stefano Negro, Francesco Ravanini, Fabio Sailis and Takato Yoshimura for useful discussions. 
 Riccardo Travaglino thanks the Department of Mathematics at City, University of London for hospitality during a three month visit in 2023 when this project was initiated. Michele Mazzoni is grateful for funding under the EPSRC Mathematical Sciences Doctoral Training Partnership EP/W524104/1 and to the organisers of the workshop ``Emergent Hydrodynamics of Integrable Systems and Soliton Gases", held in Luminy (France) in November 2023 for the opportunity to present a talk on this work.  Michele and Riccardo also thank the organisers of the ``SFT 2024 Lectures on Statistical Field Theories" (February 2024) at GGI Florence (Italy), where final work on this draft was carried out.

\appendix

\section{Multi-Particle Spectra}
\label{appA}
 For systems with several particle types labeled by $a=1,\ldots,N$ and diagonal scattering, the TBA equations take the form:
\beq
\varepsilon^\alpha_a = \nu_a - \sum_b \varphi^0_{ab} * L^\alpha_b(\theta) + \sum_b \alpha m_a m_b (\cosh*L^\alpha_b)(\theta).
\eeq
The dressing equation  \eqref{dressingT} can be easily generalised to this situation. We rewrite it in matrix form by introducing the matrix of integral operators $\mathbf{T} \mathbf{n}^\alpha$, with components $\mathbf{T}_{ab} n_b^\alpha$ (the index $b$ is not summed over). In this way $[({\bf T} {\bf n}^\alpha){\bf h}^{\rm dr}]_a(\vartheta) =\sum_b(\varphi^0_{ab} * (n^\alpha_b\, h_b^{\rm dr}))(\vartheta)$, and equations (\ref{dressingT}) and (\ref{eq:TildeDressingEquation}) still hold: 
\beq
    {\bf h}^{\rm dr,\alpha}_s(\vartheta) = (1 - \mathbf{T} {\bf n}^\alpha)^{-1} ( {\bf h}_s(\vartheta) - \alpha {\texttt q}_s^\alpha {\bf E}(\vartheta) + \alpha {\texttt j}_s^\alpha {\bf P}(\vartheta))\,,
    \label{dressingT2}
\eeq
that is, indicating with a tilde the action of the integral operator $(1 - \mathbf{T} \mathbf{n}^\alpha)^{-1}$, we have
\beq
    {\bf h}^{\rm dr, \alpha}_s(\vartheta) = \tilde{\bf h}_s^\alpha(\vartheta) - \alpha {\texttt q}_s^\alpha \Tilde{\bf E}^\alpha(\vartheta) + \alpha {\texttt j}_s^\alpha \Tilde{\bf P}^\alpha(\vartheta)\,.
    \label{eq:TildeDressingEquationMultiP}
\eeq 
Notice that now  ${\bf h}^{\rm dr,\alpha}_s(\vartheta), {\bf h}_s(\vartheta), {\bf n}^\alpha(\vartheta), {\bf E}(\vartheta), {\bf P}(\vartheta)$, together with their tilded versions, are $n$-component vectors (hence the bold font). On the other hand, ${\texttt q}_s^\alpha$ and ${\texttt j}_s^\alpha$ are the total averages, that is scalars which already include the sum of contributions from all particle species. The inversion of a matrix of integral operators is delicate and has to be dealt with carefully. The series expansion of the inverse operator, which we introduced in Section~\ref{sec:mainresults}, must converge for physical reasons, otherwise the dressing operation would be ill-defined. This is however ensured here by the fact that the kernel operator involved is that of a known IQFT, and these always display good convergence properties (they are typically exponentially decaying functions for large $|\vartheta|$). 

The discussion of Section \ref{sec:mainresults} also follows through regarding the massless and the equilibrium limit. In both cases, the dressing operator has the effect of redefining the inverse temperature(s) in the underformed theory. Recall that we denoted quantities at inverse temperature $\hat{\beta}$ (or $\hat{\beta}_{R,L}$ in the partitioning protocol) with a hat, so tildes are replaced by hats everywhere and:
\begin{equation}
    {\bf h}^{\rm dr,\alpha}_s(\vartheta) = \hat{{\bf h}}^{\rm dr,0}_s(\vartheta) - \alpha {\texttt q}_s \hat{{\bf E}}^{\rm dr} + \alpha {\texttt j}_s \hat{{\bf P}}^{\rm dr} \text{   as  } m \to 0\,.
\end{equation}
The total average currents and densities are given again by (\ref{eq:expressionnoice}) in the massive case and by (\ref{34}) in the massless and equilibrium limits. The solution will be exactly the same as that given previously. This shows in particular the universality of our results, with (\ref{critical}) valid for any $\TTb$-perturbed CFT, (\ref{eq:TildeSystem}) valid in massive, interacting, out-of-equilibrium theories and (\ref{freef}) valid at equilibrium for massive and massless theories. {\iffalse\color{red}\fi This applies in particular to the famous ADE models described in \cite{ZamolodchikovADE} and \cite{dynkin}. An obvious generalisation would be to consider non-diagonal theories, for example those described by magnonic TBA equations \cite{Quattrini:1993sm, RAVANINI199273}. Such theories are characterised by TBA equations exhibiting both massive (physical) and massless (magnonic) excitations. A major complication comes the fact that the $T\overline{T}$-deformation acts differently on the physical quasiparticles and on the magnons, thus leading to rather intricate TBA equations for the perturbed theory. The situation is even more complicated for models in which the TBA approach in the unperturbed theory is challenging in itself, such as the Sine-Gordon model. A general approach for such theories is still missing, although the expectation is that our results should be valid for any kind of scattering, as is suggested by looking at the theory directly from the CFT limit, as shown in appendix \ref{appC}.}


\section{Generalised Deformations}
\label{appB}
It is possible, with some caveats, to extend the discussions to the case of generalised $\TTb$ deformations. The TBA for this situation was studied in detail in \cite{Hernandez-Chifflet:2019sua}. Here, we will restrict ourselves to the special case of a driving term $\nu(\vartheta)=m^s\cosh{s\vartheta}$ and $S$-matrix deformation given by $e^{-i \delta(\vartheta)}$ with\footnote{The factor $\frac{1}{s}$ is included for convenience, but it can always be absorbed by a redefinition of $\alpha$. As discussed in \cite{Hernandez-Chifflet:2019sua}, there may be convergence issues if the parameter $s$ in the driving term is different from that of the deformation.} $\delta(\vartheta)=\frac{m^{2s}\alpha \sinh(s\theta)}{s}$.  For simplicity, we will focus on the free fermion theory, even if results in the massless limit hold more generally.  Following the same kind of derivation as presented in Section \ref{sec:tbadressing} we find that, at equilibrium
\begin{eqnarray}
\label{generalised_TT_TBA_eq}
    \varepsilon^\alpha(\vartheta) 
    = m^s\beta^s \cosh(s\vartheta) +  m^{2s} \alpha\cosh(s\vartheta) \int \frac{d\vartheta'}{2\pi}\cosh(s\vartheta')L(\vartheta')\,.
\end{eqnarray}
We can then define a generalisation of the free TBA energy that we considered before:
\begin{equation}
    {E}_s^\alpha := -\frac{m^s}{2\pi} \int d\vartheta \cosh(s\vartheta) L(\vartheta)\,.
\end{equation}
This object is interpreted as the analogue of $E_0^\alpha$ for higher spin charges, and we will refer to it as a generalised energy. Introducing this quantity, the TBA equation becomes:
\begin{equation}
    \varepsilon^\alpha(\vartheta) = m^s\beta^s \cosh(s\vartheta) -\alpha m^s {E}_s^\alpha \cosh(s\vartheta).
\end{equation}
From this we find a self consistent equation for the generalised energy:
\begin{equation}
    {E}_s^\alpha=-\frac{m^s}{2\pi}\int \cosh (s\vartheta)\log(1+e^{-m^s(\beta^s-\alpha {E}_s^\alpha)m^s \cosh(s\vartheta)})\,. 
\end{equation}
This is exactly the same formula as (\ref{ole}) except for a redefinition of the coefficient of $\cosh(s\vartheta)$, so we can carry out the same type of computation based on 
the small $m$ expansion of the Bessel functions, obtaining:
\begin{eqnarray*}
     {E}^\alpha_s = -\frac{\pi c}{6 s (\beta^s-{\alpha}{E}_s^\alpha)} \qquad \mathrm{for}\quad m\rightarrow 0\,.
\end{eqnarray*}
This gives
\begin{equation}
  {E}^\alpha_s =  \frac{\beta^s}{2}\left(1-\sqrt{1+\frac{2\pi c \alpha }{3 s \beta^{2s}}} \right)\,,
\end{equation}
where again we chose the solution with negative sign in front of the square root. 
 Therefore, defining the (generalised) modified inverse temperature as $\hat{\beta}^s= \beta^s-\alpha {E}_s^\alpha$, we obtain 
\begin{equation}
    \hat{\beta}^s = \frac{\beta^s}{2}\left(1+\sqrt{1+\frac{2\pi c \alpha }{3 s \beta^{2s}}} \right)\,.
    \label{powers}
\end{equation}
In terms of $\hat{\beta}^s$ the equilibrium TBA equation \eqref{generalised_TT_TBA_eq} reads:
\begin{equation}
\label{eq:finalgeneralisedtba}
    \varepsilon^\alpha(\vartheta) = \hat{\beta}^s m^s \cosh(s\vartheta)\,,
\end{equation}
while the TBA for interacting theories with a generalised $\TTb$ perturbation is obtained by introducing a non vanishing kernel $\varphi^0$ in \eqref{generalised_TT_TBA_eq}. It is also possible to carry out a similar analysis in a multi-particle theory with the replacement $\alpha m^{2s} \mapsto \alpha m_i^s m_j^s$.

Since the modified inverse temperature has no mass dependence, the CFT limits of \eqref{eq:finalgeneralisedtba} for right and left movers follow straightforwardly:
\begin{equation}
   \varepsilon^\alpha_R(\vartheta) =  \frac{M^s  \hat{\beta}^s}{2} e^{s {\vartheta}}\,, \quad \varepsilon^\alpha_L(\vartheta) =  \frac{M^s  \hat{\beta}^s}{2} e^{-s {\vartheta}}\,,
\end{equation}
where as usual $M := m e^{\vartheta_0}$, with $\vartheta_0$ the divergent part of the rapidity. Hence the massless TBA equations become in this case:
\begin{equation}
\label{eq:generalisedmasslesstba}
    \varepsilon^\alpha_{\pm}(\vartheta) = \frac{M^s \beta^s }{2} e^{\pm s\vartheta} -\frac{\alpha_s M^{s}}{2} {E}_s^\alpha e^{\pm s\vartheta} \pm \frac{\alpha_s M^s}{2}{P}_s^\alpha e^{\pm s\vartheta}\,, 
\end{equation}
where we included the term proportional to $P_s^\alpha$, which is the generalisation of $P_0^\alpha$ as defined in (\ref{EP}) and vanishes if the system is at equilibrium.

\subsection{Averages}
 Here, we present a computation of the average charge densities in the presence of a generalised $\TTb$ deformation of the free fermion at equilibrium. Consider a deformation that contains a spin-$s$ term, so that the TBA equation is \eqref{eq:finalgeneralisedtba}. If we are interested in average densities and currents of a spin-$s'$ charge, then the dressing relation is
\begin{equation}
    h_{s'}^{\rm dr, \alpha}(\vartheta)= h_{s'}(\vartheta) -\alpha m^{2s} \cosh(s\vartheta) \int_{-\infty}^{\infty} \frac{d\vartheta'}{2\pi}\cosh(s\vartheta') n^{\alpha}(\vartheta') h_{s'}^{\rm dr,\alpha}(\vartheta')\,,
\end{equation}
as it simply follows from the definition \eqref{dressing1}. The energy dressing is particularly simple and takes the form 
\begin{equation}
    E^{\rm dr,\alpha}(\vartheta) = E(\vartheta) - \alpha {\texttt q}_s^\alpha h_s(\vartheta)\,.
\end{equation}
For $s=s'$ we can compute
\beq
    {\texttt q}_s^\alpha = \frac{1}{2\pi} \int d\vartheta E^{\rm dr,\alpha}(\vartheta) n^\alpha(\vartheta) h_s(\vartheta) =\frac{m^s}{2\pi} \int d\vartheta (m\cosh{\vartheta} - \alpha m^{s} {\texttt q}_s^\alpha\cosh{s\vartheta} ) n^\alpha(\vartheta) \cosh(s\vartheta)\,,
    \eeq 
and since $n^\alpha(\vartheta)=(1+e^{\varepsilon^\alpha(\vartheta)})^{-1}$ we can again expand in terms of Bessel functions using (\ref{49}) and (\ref{50}) to arrive at
\beqa
{\texttt q}_s^\alpha &=& \sum_{n=1}^\infty (-1)^{n+1} \frac{m^{s+1}}{\pi s } \left(K_{1+1/s}(n m^s \hat{\beta}^s)- \frac{1}{s n m^s \hat{\beta}^s} K_{1/s}(n m^s \hat{\beta}^s)\right)\nonumber\\
&& - \sum_{n=1}^\infty (-1)^{n+1} \frac{\alpha m^{2s} {\texttt q}_s}{\pi s }   \left(K_2(n m^s \hat{\beta}^s)- \frac{1}{n m^s \hat{\beta}^s} K_{1}(n m^s\hat{\beta}^s)\right)\,,
\eeqa
which, up to a factor $\frac{1}{s}$ coming from a rescaling of the rapidity, is identical to \eqref{ciao} up with the substitutions $s \to \frac{1}{s}$, $m\hat{\beta}\mapsto m^s \hat{\beta}^s$ and $m^2 \alpha \mapsto m^{2s} \alpha$. Hence, for small $m$: 
\beq 
{\texttt q}_s^\alpha=\frac{\Gamma(\frac{1}{s}+1) \zeta(\frac{1}{s}+1) (2^{\frac{1}{s}}-1)}{2\pi \hat{\beta}^{s+1}
\left(s+\frac{\alpha \pi}{12\hat{\beta}^{2s}}\right)} \,,
\label{chargegeneralised}
\eeq 
an expression which non depends on the dimensionless scale $\alpha/\hat{\beta}^{2s}$. By following the same procedure we can compute the averages of charge densities of arbitrary spin $k$, obtaining in the small $m$ limit:
\begin{equation}
\label{eq:complicatedstuff}
       {\texttt q}_{k}^\alpha = {\texttt q}_{k,s}^0 -\alpha {\texttt q}_s^\alpha \frac{\Gamma(1+\frac{k}{s}) \zeta(1+\frac{k}{s})}{2\pi s \hat{\beta}^{s+k}}(2^{k/s}-1)\,.
\end{equation}
In the equation above, ${\texttt q}_{k,s}^0$ is an $\alpha$-independent term which takes the form
\begin{equation}
 {\texttt q}_{k,s}^0 = \frac{\Gamma(\frac{k+1}{s}) \zeta(\frac{k+1}{s})}{2\pi s \hat{\beta}^{1+k}}(2^{\frac{k+1}{s}-1}-1)\,,
\end{equation}
while ${\texttt q}_s^\alpha$ is given by \eqref{chargegeneralised}. Therefore, average charge densities at equilibrium can be obtained in a rather straightforward way even for generalised deformations.

\section{CFT Derivation}
\label{appC}
\begin{figure}
    \centering
    \includegraphics[width=\textwidth]{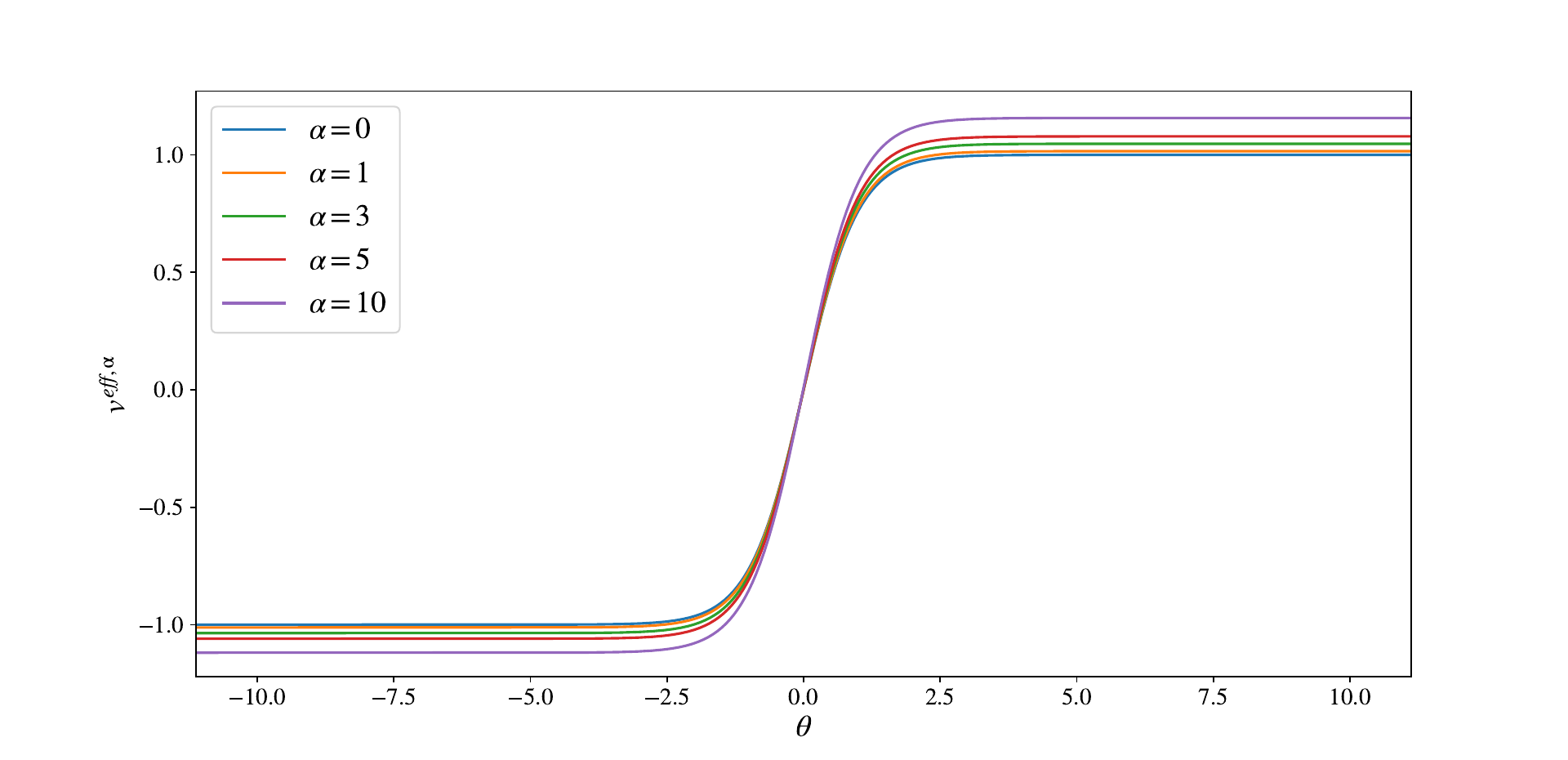}
    \caption{Behaviour of the effective velocity for a free fermionic theory for different values of $\alpha$ at fixed finite $m$ and fixed temperatures. The velocities \eqref{eq:massless_v_eff} of the right and left movers are simply the limits for $\theta \to \pm \infty$ of the plotted $v^{\rm eff, \alpha}$. The superluminal behaviour is evident from the fact that for $\alpha \neq 0$ the asymptotic values are above 1 (below -1).}
    \label{fig:veff}
\end{figure}

In the main text we obtained results for massless theories starting from massive perturbed TBA and then taking the $m\to 0$ limit. The advantage of this approach is that it allows us to understand which quantities can be found only in the perturbed CFT case and which ones can be determined also in the massive case, hence with greater generality. Moreover, the approach is useful because the TBA formulation of CFTs is in general less transparent than the massive one. Nonetheless, it is possible to find massless results starting from the TBA equations of the perturbed CFT as given by \eqref{eq:perturbedcft}. From this equation we can find the dressed eigenvalues, which are different for right and left movers:
\begin{equation}
    (h_s^{\pm})^{\rm dr, \alpha}(\vartheta)= h_s^{\pm}(\vartheta) - \alpha h^{\pm}_1(\vartheta) (\texttt{q}_s^\alpha \mp {\texttt j}_s^\alpha) + (\varphi^0 * n^{\pm,\alpha} (h_s^{\pm})^{\rm dr})(\vartheta)\,,
\end{equation}
where the one-particle eigenvalues of the massless excitations are simply
\begin{equation}
    h_s^{\pm}(\vartheta)=\frac{M^s}{2}e^{\pm s\vartheta}.
\end{equation}
As before we can invert the dressing operation:
\begin{equation}
    (h_s^{\pm})^{\rm dr, \alpha}(\vartheta)= (1+\mathbf{T}n^{\pm})^{-1} \left(h_s^{\pm} - \alpha h^{\pm}_1 ({\texttt q}_s^\alpha \mp {\texttt j}_s^\alpha)\right).
\end{equation}
In the massless case it is clear that the occupation functions $n^{\pm, \alpha}$ are the same as those of the unperturbed theory up to a redefinition of temperature. This leads to much simpler expressions for the effective velocities of the right and left movers. Indeed, from
\begin{equation}
    (h_1^{\pm})^{\rm dr,\alpha}(\vartheta) = (1-\alpha ({\texttt q}_s^\alpha \mp {\texttt j}_s^\alpha) )(1+\mathbf{T}n^{\pm})^{-1}  h_1^{\pm}(\vartheta),
\end{equation}
it follows that the effective velocities of the two species do not depend on the rapidity:
\begin{equation}
\label{eq:massless_v_eff}
    (v^{\rm eff,\alpha})^{\pm} =  \pm \frac{ 1+\alpha({\texttt q}_1^\alpha \mp {\texttt j}_1^\alpha)}{1-\alpha({\texttt q}_1^\alpha \mp {\texttt j}_1^\alpha)}=\pm \left( 1 + \frac{2 \alpha({\texttt q}_1^\alpha \mp {\texttt j}_1^\alpha)}{1-\alpha({\texttt q}_1^\alpha \mp {\texttt j}_1^\alpha)}\right)\,.
\end{equation}
The fact that the effective velocities are simply shifted by fixed constant terms was already found in \cite{MedenjakLong}. The quantity in brackets in the above expression is usually larger than one in absolute value, hence giving rise to superluminal effects, as manifest from figure \ref{fig:veff}. 
 In any case, the solution of the partitioning protocol is not modified, and the equality which relates the dressed quantities of the perturbed theory to the dressed quantities of the unperturbed theory is
\begin{equation}
    (h_s^{\pm})^{\rm dr, \alpha}(\vartheta)=(\hat{h}_s^{\pm})^{\rm dr, \alpha}(\vartheta) - \alpha ({\texttt q}_s^\alpha \mp {\texttt j}_s^\alpha) (\hat{h}_1^{\pm})^{\rm dr, \alpha}(\vartheta)\,.
\end{equation}
The evaluation of the currents and densities has therefore to take into account the sum over the right and the left movers. 
Observing that $\hat{\texttt q}_1^{+}+\hat{\texttt q}_1^{-}=\hat{\texttt q}_1^0$, and  $\hat{\texttt q}_1^{+}-\hat{\texttt q}_1^{-}=\hat{\texttt j}_1^0$, we recover the result for massive theories:
\begin{equation}
     {\texttt q}_s^\alpha = \hat{\texttt q}_s^0 -\alpha \hat{\texttt q}_1^0 {\texttt q}_s^\alpha +\alpha \hat{\texttt \j}_1^0 {\texttt j}_s^\alpha\,. 
\end{equation}
An identical discussion can be also performed for the current, leading to a system of equations which is identical to \eqref{eq:expressionnoice} but is directly evaluated in the massless limit. 
Therefore, the discussion can be simply repeated using the results of \cite{MazzoniUnpublished} to obtain the same expressions \eqref{eq:finalcurrent} and \eqref{eq:finalcharge} directly in the conformal limit.

\section{$\hat{\beta}$ as an Explicit Function of ${\beta}$: the free Fermion Case}
\label{appD}
Except for the general properties discussed above, for most theories it is not possible to find a closed-form expression for the function $c^\alpha(r,r')$ beyond the critical point. Once more, the massive free fermion provides an exception to this rule, in that the $c$-function admits a perturbative expansion in terms of $\hat{r}=m\hat{\beta}$. This allows us to find a perturbative expansion of $\hat{\beta}$ in terms of ${\beta}$ which holds in the massive regime and exhibits some interesting mathematical features. 

Consider again equation \eqref{eq:groundstateenergy}. This can be rewritten in terms of Bessel functions, as shown in the main text:
\beqa 
E_0^\alpha =\frac{m}{2\pi} \sum_{n=1}^\infty \frac{(-1)^n}{n}\int_{-\infty}^\infty \cosh\vartheta e^{-n  \hat{r}\cosh\vartheta} d\vartheta= \frac{m}{\pi} \sum_{n=1}^\infty \frac{(-1)^n}{n} K_1(n\hat{r})\,.
\eeqa 
In \cite{KLASSEN1991635} a solution to this equation for $\alpha=0$ was presented, in the sense that the free energy, or rather the scaling function, was obtained as a perturbative expansion in the parameter $r$. By using the same formula, we can expand in $\hat{r}$ the right-hand side of the previous equation:  
\beqa
c^0(\hat{r})
= \frac{1}{2}-\frac{3\hat{r}^2}{2\pi^2} \left[\ln\frac{\hat{r}}{\pi} - \frac{1}{2}+ \gamma_E \right]  - \frac{6}{\pi}
\sum_{n=1}^{\infty}\left( \sqrt{(2n-1)^2\pi^2 + \hat{r}^2 } - (2n-1)\pi -\frac{\hat{r}^2}{2(2n-1)\pi}  \right),
\label{eq:c-function}
\eeqa
where $\gamma_E=0.577216...$ is the Euler-Mascheroni constant.
The formula can be slightly simplified by expanding the square root as:
\beqa
     \sqrt{(2n-1)^2\pi^2 + \hat{r}^2 } &=& (2n-1)\pi \sum_{k=0}^{\infty} {1/2\choose k} \left(\frac{\hat{r}^2}{(2n-1)^2\pi^2} \right)^k \nonumber \\
     &=& (2n-1)\pi +  \frac{\hat{r}^2}{2(2n-1)\pi} + \sum_{k=2}^{\infty} {1/2\choose k} \left(\frac{\hat{r}^2}{(2n-1)^2\pi^2} \right)^k\,,
\eeqa
where the first two terms cancel off the last two terms in the sum \eqref{eq:c-function}. Therefore, the scaling function reads:
\begin{equation}
    c^0(\hat{r}) = \frac{1}{2}-\frac{3\hat{r}^2}{2\pi^2} \left[\ln\frac{\hat{r}}{\pi} - \frac{1}{2} + \gamma_E \right] + 6 \sum_{k=2}^{\infty} {1/2\choose k} \frac{\hat{r}^{2k}}{\pi^{2k}} \sum_{n=1}^{\infty}\frac{1}{(2n-1)^{2k-1}}\,.
\end{equation}
The sum in $n$ is given in terms of Riemann's zeta function
\beq
\sum_{n=1}^{\infty}\frac{1}{(2n-1)^{p}}= (1-2^{-p})\zeta(p)\,,
\eeq
yielding the more compact expression:
\beq
  c^0(\hat{r}) = \frac{1}{2}-\frac{3\hat{r}^2}{2\pi^2} \left[\ln\frac{\hat{r}}{\pi} - \frac{1}{2}  + \gamma_E \right] + 6 \sum_{k=2}^{\infty} {1/2\choose k} \frac{\hat{r}^{2k}}{\pi^{2k}} (1-2^{1-2k})\zeta(2k-1)\,.
\eeq
This formula allows us to expand $m^{-1}E_0^\alpha = -\frac{\pi c^0(\hat{r})}{6\hat{r}}$ as a perturbative series in $\hat{r}$. Furthermore, by using $ m^{-1} E_0^\alpha = \frac{r - \hat{r}}{r'}$ we can find an explicit expansion of $r$ in terms of $\hat{r}$. We obtain
\beq
\frac{r}{r'} = \frac{\hat{r}}{r'} -\frac{\pi}{12\hat{r}} + \frac{\hat{r}}{4\pi}\left[\ln{\hat{r}}+\chi\right] -  \sum_{k=2}^{\infty} {1/2\choose k} \frac{\hat{r}^{2k-1}}{\pi^{2k-1}} (1-2^{1-2k})\zeta(2k-1)\,, 
\label{eq:betaasfuncitonofbetahat}
\eeq
where $\chi = -\frac{1}{2}-\ln\pi+\gamma_E$. The equation above can be solved (at least numerically and within the radius of convergence of the series) to find the value of $\hat{\beta}$ at all orders for a massive free fermion

\subsection{Corrections Near Criticality}
Let us consider the leading corrections for small $m$. Then, we can approximate equation \eqref{eq:betaasfuncitonofbetahat} to:
\begin{equation}
\frac{r}{r'} = \frac{\hat{r}}{r'} -\frac{\pi}{12\hat{r}} + \frac{\hat{r}}{4\pi}\ln{\hat{r}}+\cdots 
\label{eq:lowestordercorrectionlog}
\end{equation}
where we neglected the $\chi$ term, since it results in a next-to-leading correction.
We can start by finding the effective temperature at $r=\beta=0$. Exponentiating the truncated equation \eqref{eq:lowestordercorrectionlog} at $r= 0$ we obtain:
\begin{equation}
\label{truncated}
    \hat{r} \exp\left(-\frac{\pi^2}{3\hat{r}^2 }\right)=\exp\left(-\frac{4\pi}{r'}\right).
\end{equation}
 This equation can be solved exactly using Lambert's $W$ function \cite{LambertFunction}, which is defined by the equation 
\begin{equation}
    W(x)e^{W(x)}=x\,.
    \label{eq:lambert}
\end{equation}
Indeed, by defining $t=\hat{r}^{-2}$ and then squaring and inverting both sides of \eqref{truncated} we have:
\begin{eqnarray*}
     \frac{2\pi^2}{3}t \exp{\left({\frac{2\pi^2}{3}t}\right)} &=&\frac{2\pi^2}{3} e^{\frac{8\pi}{r'}},
\end{eqnarray*}
which can be solved immediately using the definition \eqref{eq:lambert}:
\begin{equation}
\hat{r}_0 := \frac{\sqrt{\frac{2\pi^2}{3}} }{\sqrt{W\left(\frac{2\pi^2}{3}e^{\frac{8\pi}{r'}}\right )}} \quad \mathrm{for}\quad m\ll 1\quad \mathrm{and}
\quad \beta=0\,.
\end{equation}
Note that, by using the relation $e^{W(x)}=\frac{x}{W(x)}$, the previous expression can be rewritten as:
\begin{equation}
\label{eq:exponentiallambert}
     \hat{r}_0 = \frac{\exp{\frac{1}{2}W(\eta)}}{\exp(\frac{4\pi}{r'})} = \sqrt{\frac{2\pi^2}{3\eta}}\exp{\frac{1}{2}W(\eta)},
\end{equation}
where we introduced the parameter $\eta = \frac{2\pi^2}{3}\exp(\frac{8\pi}{r'})$. This expression will be useful in the following.
For $m\to 0$, $\hat{r}_0$ is finite, since the Lambert function behaves asymptotically as the logarithm of its argument. In particular, the limit evaluates to:
\begin{equation}
    \lim_{m\to 0} \hat{r}_0 = \sqrt{\frac{\pi r}{6}}\,,
\end{equation}
which corresponds precisely to the result obtained in the $\beta \to 0$ limit of equation \eqref{eq:defBeta}, with $c=1/2$. If $\beta \neq 0$, the solution in terms of Lambert's function is not exact, but if we assume $\beta \ll \sqrt{\alpha}$, so that $\hat{\beta}_0\approx \sqrt{\frac{\pi \alpha}{6}} \gg \beta$, then we can proceed further in our derivation. 
The procedure is analogous to the $\beta=0$ case, but with the introduction of an extra term:
\begin{eqnarray}
    \hat{r}\exp{\left( -\frac{\pi^2}{3\hat{r}^2}\left(1+\frac{12r\hat{r}}{ \pi r'}\right)\right)}=  e^{-\frac{4\pi}{r'}}\,.
    \label{neweq}
\end{eqnarray}
Unfortunately, this equation is still not solvable in terms of Lambert's function, because of the extra term in the exponential in the left hand side. Although generalisations of Lambert's function exist, we have found no generalisation that solves an equation of the type $W_a(x) e^{W_a(x)(W_a(x)+a)}=x$. Here, we assume the existence of such a function $W_a(x)$, in terms of which equation (\ref{neweq}) takes the form
\begin{equation}
    \frac{\pi}{\sqrt{3} \hat{r}}= {W}_{\frac{4 \sqrt{3} r}{ r'}}\left(\frac{\pi}{\sqrt{3}}e^{\frac{4\pi}{r'}}\right)=W_{\frac{4 \sqrt{3} r}{ r'}}\left(\sqrt{\frac{\eta}{2}}\right)\,, 
\end{equation}
in the limit of small mass and at any inverse temperature $\beta$. 

In the absence of more information about the function $W_a(x)$ we can proceed by assuming that $r$ is small and further approximating $\hat{r}\approx \hat{r}_0$ in the exponential (\ref{neweq}). 
We can then write:
\begin{equation}
\label{NLO}
      \hat{r}_1\exp{\left(- \frac{\pi^2}{3\hat{r}_1^2}\left(1+\frac{12}{\pi r'}r\hat{r}_0\right)\right)} = e^{-\frac{4\pi}{r'}}\,,
\end{equation}
where we now use the notation $\hat{r}_1$ to indicate that this expression provides a next-to-leading order approximation (in $\beta$) of $\hat{r}$. This procedure can then be generalised to higher orders, as we see below. As before, equation \eqref{NLO} can be solved exactly, this time yielding:
\begin{equation}
    \hat{r}_1=\frac{\sqrt{\frac{2\pi^2}{3}} \sqrt{1+\frac{12}{\pi r'}\beta\hat{r}_0}}{m\sqrt{W\left(\frac{2\pi^2}{3}\left(1+\frac{12}{\pi r'}r \hat{r}_0\right) e^{\frac{8\pi}{r'}}\right)}}.
    \label{eq:logcorrectionbetahat}
\end{equation}
For consistency, we can check that expression \eqref{eq:logcorrectionbetahat} leads to the correct conformal limit, which we read from \eqref{confdef}. Indeed, we have
\begin{equation}
    \lim_{m \to 0}\hat{r}_1 = \sqrt{\frac{\pi r'}{12}\left(1+r \sqrt{\frac{12}{\pi r'}}\right)} \approx \sqrt{\frac{\pi r'}{12}}+ \frac{r}{2},
\end{equation}
which is the correct result. 
Proceeding as before, from $e^{W(x)}=\frac{x}{W(x)}$ we can cast expression \eqref{eq:logcorrectionbetahat} in a form similar to \eqref{eq:exponentiallambert}:
\begin{equation}
    \hat{r}_1 = \sqrt{\frac{2\pi^2}{3\eta}} \exp\left(\frac{1}{2} W(\eta (1+ \mathcal{K}\exp{(\frac{1}{2}W(\eta))}))\right), 
\end{equation}
where the new parameter is $\mathcal{K}=\frac{12 r }{\pi r'} \sqrt{\frac{2\pi^2}{3\eta }}$. This suggests that the complete solution will be given by infinitely many ``nested" Lambert functions, as can be seen by iterating procedure above. In general, we have
\begin{equation}
    \hat{r}_{i+1} = \frac{\sqrt{\frac{2\pi^2}{3}} \sqrt{1+\frac{12}{\pi r'} r \hat{r}_i}}{\sqrt{W\left(\frac{3\pi^2}{2}\left(1+\frac{12}{\pi r'}r\hat{r}_i\right) e^{\frac{8\pi}{r'}}\right)}}\,,
\end{equation}
and the solution has structure
\begin{equation}
\label{eq:beautifulbetahat}
    \hat{r} = \sqrt{\frac{2\pi^2}{3\eta }} \exp\left(\frac{1}{2}(W(\eta(1+\mathcal{K}\exp(\frac{1}{2}W(\eta(1+\mathcal{K}\exp(\frac{1}{2}W(...))))\right)\,.
\end{equation}
This is the exact expression of $\hat{r}$ for any value of $r$, with small mass. Truncating at the $n$-th nested Lambert function and taking the massless limit gives the expansion of $\hat{r}$ in \eqref{eq:defBeta} at order $\mathcal{O}(r^{n-1})$ . An extension of \eqref{eq:beautifulbetahat} to include higher order terms in the mass remains elusive at this point. From our definition of the function $W_a(x)$, the previous expression also implies the functional relation
\beq 
{W}_{\frac{4 \sqrt{3} r}{ r'}}\left(\sqrt{\frac{\eta}{2}}\right)=\sqrt{\frac{\eta}{2\pi}} \exp\left(-\frac{1}{2}(W(\eta(1+\mathcal{K}\exp(\frac{1}{2}W(\eta(1+\mathcal{K}\exp(\frac{1}{2}W(...))))\right)\,.
\eeq 

\printbibliography
\end{document}